\documentstyle[eqsecnum,pre,preprint,aps,epsfig]{revtex}

\tightenlines

\newcommand{\bq}{\begin{equation}}
\newcommand{\eq}{\end{equation}}
\newcommand{\bqa}{\begin{eqnarray}}
\newcommand{\eqa}{\end{eqnarray}}

\def\prep#1#2#3{ Phys. Rep. ${\bf{#1}}$ (#2) #3}
\def\etal{{\it et.al.\/}}

\begin{document}

\draft
\preprint{PM/01-16, hep-ph/0103335, corrected version}

\title{The role of universal and non universal Sudakov logarithms\\
 in four fermion processes at TeV energies:\\
 the one-loop approximation
revisited.\footnote{Partially
supported by EU contract  HPRN-CT-2000-00149.}}

\author{M. Beccaria$^{a,b}$, F.M.
Renard$^c$ and C. Verzegnassi$^{d,e}$ \\
\vspace{0.4cm} 
}

\address{
$^a$Dipartimento di Fisica, Universit\`a di 
Lecce \\
Via Arnesano, 73100 Lecce, Italy.\\
\vspace{0.2cm}  
$^b$INFN, Sezione di Lecce.\\
\vspace{0.2cm} 
$^c$ Physique
Math\'{e}matique et Th\'{e}orique, UMR 5825\\
Universit\'{e} Montpellier
II,  F-34095 Montpellier Cedex 5.\hspace{2.2cm}\\
\vspace{0.2cm} 
$^d$
Dipartimento di Fisica Teorica, Universit\`a di Trieste, \\
Strada Costiera
 14, Miramare (Trieste) \\
\vspace{0.2cm} 
$^e$ INFN, Sezione di Trieste\\
}

\maketitle

\begin{abstract}

We consider the separate effects on four fermion processes, in the TeV
energy range, produced at one loop by Sudakov logarithms of universal
and not universal kind, working in the 't Hooft $\xi=1$ gauge. Summing
the various vertex and box contributions allows to isolate two quite
different terms.The first one is a combination of vertex
and box quadratic and linear logarithms that are universal 
and independent of the scattering angle
$\theta$. The second one is $\theta$-dependent, not universal, 
linearly logarithmic and only produced by weak boxes. We show that 
for several observables, measurable at future linear $e^+e^-$ 
colliders (LC, CLIC), the role of the latter term is dominant
and we discuss the implications of this fact for what concerns 
the reliability of a one-loop approximation.

\end{abstract}


\section{Introduction.} 

The fact that double logarithms of the c.m. energy $\sqrt{q^2}$
of "Sudakov-type" \cite{Sudakov} will affect 
the asymptotic behaviour of
the weak component of four-fermion processes at one-loop has been known
since a decade \cite{KDD}. Only recently, though, it has been realized
that this effect could be already relevant for c.m. energies in the
${\rm TeV}$
range \cite{rising}, which is the case of the next generation of linear
$e^+e^-$ colliders at $500~{\rm GeV}$ (LC)\cite{LC} and $3~{\rm TeV}$
(CLIC)\cite{CLIC}. In fact, a first explicit calculation of the double
logarithms effects \cite{CC} showed that their size at ${\rm TeV}$ energies
might well cross the relative ten percent value. In a following paper
\cite{log}, the extra effects at one loop of the subleading linear
logarithms on a class of observables (cross sections, asymmetries) were
thoroughly computed, showing the existence of a certain cancellation
between the leading and the subleading terms in a "pre-asymptotic"
energy region, whose details strongly depend on the considered final
state and observable \cite{mt,DenPo}.
As a general conclusion, it was stressed in
Refs.\cite{CC,log,mt,DenPo} 
that the validity of a one loop approximation 
for four-fermion processes in
the ${\rm TeV}$ region was not obvious, and seemed also to depend strongly on
the particular chosen observable.\par
To get rid of low convergence rate in the perturbative expansion,
several papers were written in very recent times with the specific aim
of resumming the dangerous leading components of the Sudakov logarithms
to all orders \cite{CCF}, or at least to compute them at two loops
\cite{MBH}. Since, as already noticed in Ref.\cite{log}, the role of
the subleading logarithms is crucial already
at the one loop level, the next effort was
that of trying to resum to all orders both leading and subleading
Sudakov logarithms. This task has been performed in at least two recent
papers \cite{KPS,Melles}. The conclusion of 
Ref.\cite{Melles} is that the set of
Sudakov logarithms can be divided into two subsets, that can be
distinguished at the level of invariant scattering amplitude since the
first subset provides a universal
contribution that is only dependent on the c.m.
energy $\sqrt{q^2}$, whilst the second, not universal one, is not only
depending on $\sqrt{q^2}$ but also on the scattering angle $\theta$.
While for the first subset precise prescriptions exist that allow to
resum to all orders the involved logarithms, for the
$\theta$-dependent part a clean resummation prescription
does not seem to exist at the moment \cite{Melles}. Should the
role of this angular component become relevant for some specific
observables, the problem of, at least, computing it to the next two
loop level might arise for the purpose of a high precision
determination of the Standard Model component of those quantities, and this
calculation does not appear to be an easy task.\par
The aim of this paper is precisely that of investigating the role of
the $\theta$-dependent non universal components of the Sudakov
logarithms, via an exhaustive analysis  of the size of their effect on
the class of experimental observables that will be measured at the next
linear electron-positron colliders, LC and CLIC. This analysis will be
performed at the one loop level, under realistic assumptions concerning
the available experimental accuracy of these machines. Our philosophy
will be that of assuming that any gauge invariant effect that is
"tolerable", given the expected experimental accuracy, 
at the one loop level,
can safely be computed in that approximation, thus avoiding the hard
problem of a two loop analysis. We shall be working in the familiar
't Hooft $\xi=1$ gauge, where all the Sudakov logarithms at one loop are
produced by either vertices or boxes. Our analysis will only consider
what we call "genuine" weak effects, due to exchanges of virtual $W's$
and $Z's$ (and, possibly, would-be Goldstone bosons for final "massive"
quarks). All other diagrams will be considered as belonging to the
"QED" component, and not discussed in this paper. For instance, we
shall not include in our analysis the boxes with one photon and one $Z$
as done in Ref.\cite{KPS}. At the one loop level, this will create very
small
numerical differences that will not change any of our conclusions,
as we will see later on.\par
Technically speaking, the paper will be organized as follows. In
Section II we shall discuss the way in which the $\theta$-independent
terms always combine in a special expression 
$U(q^2)=3\ln q^2-\ln^2 q^2$. This is precisely the
combination which was shown to exponentiate in the massless quarks
approximation \cite{Melles}. The remaining Sudakov contributions are
those of Yukawa origin ($\simeq m^2_t, m^2_b$) and the
$\theta$-dependent ones. In Section III, we discuss the different roles
of the various terms in the various experimental observables, with
special emphasis on the two LC and CLIC situations. We shall see that
there are energies and observables where the role of the
$\theta$-independent terms can be totally neglected, so that the bulk
of the
Sudakov effects at one loop is given by the corresponding angular
dependent part of the "genuine electroweak" boxes. 
In the final Section IV, we
shall discuss the validity of the one loop approximation for various
observables at various energies. In particular, we shall try to clarify
the role of some approximations that were used in our approach and also
in other, similar ones, with which we shall compare our numerical
results. This should allow us to draw a number of general conclusions. 
A short Appendix A will be devoted to an investigation of
the Sudakov effect in the special case of forward-backward asymmetries,
that seem to have a peculiar role in this case. In Appendix B
we give the general form of the polarized differential cross sections
in our theoretical scheme, from which it is easy to get
the prediction for all the considered observables.

\section{Universal and non universal Sudakov logarithms at one loop}

We shall only consider in this paper the "genuinely electroweak"
one-loop component of the invariant scattering amplitude for the
process of electron-positron annihilation into a $f\bar f$ 
fermion-antifermion pair, where the mass of the fermion $f$ will be
retained in the two cases $f=b, t$. This means that we shall only
consider those one-loop diagrams that add to the Born structure virtual
exchanges of weak particles(i.e. $W,Z,\Phi,H$, where $\Phi^{\pm,0}$ 
are the would-be
Goldstone bosons and $H$ the physical Higgs boson). 
In particular, the boxes with one photon
and
one $Z$ will be considered as belonging to the "QED" component, to be
computed separately. This is, at one loop, a mere conventional
division, that appears to us justified to the extent that such boxes
must be in any case combined with initial and final photon emission
interference in existing computational programs \cite{QEDprog}.\par
For what concerns the practical approach, we shall follow the
prescription originally proposed a few years ago \cite{Zsub} and
called "$Z$-peak subtracted". In this approach, the invariant amplitude
is decomposed in four terms, corresponding to the four independent
Lorentz structures of the process, denoted as:

\bq {\cal M}^{(1)}_{lf}(q^2,\theta)=
{\cal M}^{(1)}_{lf,\gamma\gamma}(q^2,\theta)+
{\cal M}^{(1)}_{lf,ZZ}(q^2,\theta)+
{\cal M}^{(1)}_{lf,\gamma Z}(q^2,\theta)+
{\cal M}^{(1)}_{lf,Z\gamma}(q^2,\theta)
\label{calM}\eq
\noindent
where the ($\gamma,Z$) indices denote projections of the external
Lorentz vectors and axial vectors of the process on the photon and $Z$
structures of the initial and final particles (for example ,
${\cal M}^{(1)}_{lf,\gamma\gamma}(q^2,\theta)$ 
and  ${\cal M}^{(1)}_{lf,ZZ}(q^2,\theta)$ have the same Lorentz
structure as the photon and $Z$ exchanges at Born level). The technical
features have been exhaustively discussed in previous recent references
\cite{log,mt}, to which we defer the reader for more details. 
Here we shall
only repeat that to each of the four independent Lorentz structures,
that must be by construction evidently gauge-independent, there
corresponds at one loop a certain "form factor", depending on $q^2$
(the squared c.m. energy) and $\theta$ (the c.m. scattering angle).
This consists of a precise linear combination of self-energies, vertices
and boxes that must be, consequently, gauge-independent, as one can
easily verify e.g. by following the fundamental Degrassi-Sirlin
approach \cite{DS} to which, as usually, we shall stick in this
paper.\par
Following our previous definitions, we shall call
$\tilde{\Delta}_{\alpha,lf}(q^2,\theta)$, $R_{lf}(q^2,\theta)$,
$V_{\gamma Z,lf}(q^2,\theta)$, $V_{Z\gamma,lf}(q^2,\theta)$ these four
form factors. To derive the corresponding contributions from
self-energies, vertices and boxes is straightforward once the
"projection" operations have been defined \cite{Zsub}. This allows to
derive in an easy way the related contributions to each observable
quantity of the process, starting from the definition of differential
cross sections that can be found e.g. in the Appendix B of
Ref.\cite{log}
and that we also repeat, for sake of completeness, in the Appendix B of
this paper.\par
After this short and unavoidable preliminary summary, we are now ready
to discuss the contributions of the various one-loop diagrams to the
asymptotic Sudakov logarithms, keeping in mind the fact that our
treatment is performed in the t'Hooft $\xi=1$ gauge. Quite generally,
we shall first divide the contributions into those of "universal" and
those of "non universal" type.
Following the conventionally adopted definitions \cite{Melles}, we
shall call "universal" those corrections that only depend on the
quantum numbers of the \underline{external} lines, but are otherwise
process-independent, and "not universal" the remaining terms. In
practice, the first set will contain all those contributions that are
only depending on $q^2$, and do not depend on $\theta$. These will
always come from vertices and, partially, from boxes. The second,
$\theta$-dependent set, will be provided by a certain component of the
box diagrams, both with $WW$ and with $ZZ$ exchange, although the
latter ones will be numerically much smaller in the SM.
Note that we shall consider, at least for a first approach,
production of "physical" (i.e. \underline{not chiral}) fermions,
considering observable quantities where one sums over the final spins. A
restriction to a fixed final chirality is obviously straightforward
using the rules for projecting  $\gamma^{\mu}P_L$
and $\gamma^{\mu}P_R$ terms on the photon and $Z$ Lorentz
structure\cite{Zsub}.\par
The Feynman diagrams that produce Sudakov logarithms are shown in
Fig.1. Rather than writing their explicit contributions to the
components (vertices, boxes) of the scattering amplitude, we shall
review here those to the four independent form factors that we are
using. More precisely, we find the following results:\par
a) Vertex with one $W$ (Fig.1a). This produces a universal
contribution, always proportional to the combination 

\bq
U_W(q^2)=3 ln{q^2\over M^2_W} - ln^2{q^2\over M^2_W}
\label{UW}\eq

b) Vertex with one $Z$ (Fig.1b). This generates a similar universal
contribution, proportional to the combination

\bq
U_Z(q^2)=3 ln{q^2\over M^2_Z} - ln^2{q^2\over M^2_Z}
\label{UZ}\eq

Clearly, in an asymptotic energy regime one will be fully entitled to
write 
$$U_W(q^2)\simeq U_Z(q^2)$$
if only the leading squared logarithm has to be retained. For a more
rigorous selection that also includes the subleading linear logarithms
this is, though, not correct and we shall treat separately the two
effects in our analysis.\par
c) Vertex with two $W's$ (Fig.1c). This provides a
\underline{universal} Sudakov term that is of \underline{subleading}
(linearly logarithmic) type. Apparently this breaks the feature of
producing a  $U_W(q^2)$ term like that of eq.(\ref{UW}). In fact, in the
$\xi=1$ gauge, this term is reproduced at one loop by taking into
account the additional contributions coming from boxes with two $W's$
($W$ box), as we shall now discuss.\par
d) $W$ box (Fig.1d). This produces two quite different Sudakov terms.
The first one is of leading (quadratic) logarithmic type, is
\underline{universal} and is \underline{only $q^2$ dependent}. When
added to the universal and subleading logarithmic term produced by the
vertex with two $W's$, \underline{it recomposes the combination 
$U_W(q^2)$ of Ref.~\cite{Melles}}. 
The second contribution is non universal,
$\theta$-dependent and of subleading (linear) logarithmic type. This,
in a general analysis like that of 
Ref.\cite{Melles}, \underline{would not
exponentiate}.\par
e) $Z$ boxes (Fig.1e). These only produce a non universal,
$\theta$-dependent subleading linear logarithm. Generally speaking, the
size of this term, that would also not exponentiate, 
is much smaller than that of the corresponding one
produced by the $W$ box.\par
f) Would-be Goldstone bosons and Higgs
vertices (Fig.1f). These add extra universal,
$\theta$-independent, subleading Sudakov logarithms of Yukawa type,
that only affect final $b\bar b$ and $t\bar t$ production. In the
treatment given in Ref.\cite{Melles} they are 
grouped together with the set
of $\theta$-independent logarithms $\simeq U_W(q^2)$ ,$U_Z(q^2)$ and
exponentiate cumulatively with them.\par
Our results are exposed in analytic form in the next equations, that
provide the expressions of the four gauge-independent form factors for
final states $\mu^+\mu^-$, $u\bar u$, $d\bar d$. The extra Yukawa terms
that appear must only be retained when $u=t$, $d=b$. In our
classification, we have first listed the $\theta$-independent
contributions, putting
the one $W$ and one $Z$ vertices in the first
two terms and the universal contribution produced by the combination
of the $W$ boxes with the $2W$ vertex in the third term;
$\theta$-dependent non universal $W$ and $Z$ boxes follow in the next
terms, and the universal $\theta$-independent Yukawa contributions
appear in the last term. The obtained expressions are the following
($S \equiv$ Sudakov)

\bqa
\tilde{\Delta}^{S}_{\alpha,l\mu}(q^2,\theta)&=&
{\alpha(1-v^2_e)\over32\pi s^2_Wc^2_W}U_Z(q^2)
+{\alpha\over2\pi}U_W(q^2)\nonumber\\
&&-{\alpha\over\pi}ln[{1-cos\theta\over2}]
ln{q^2\over M^2_W}+{\alpha(1-v^2_e)^2\over256\pi
s^4_Wc^4_W}ln[{1+cos\theta\over1-cos\theta}]
ln{q^2\over M^2_Z}
\label{DAl}
\eqa

\bqa
R^{S}_{l\mu}(q^2,\theta)&=&{\alpha\over4\pi s^2_W}U_W(q^2)
-{\alpha(1+3v^2_e)\over32\pi s^2_Wc^2_W}U_Z(q^2)
-{\alpha c^2_W\over2\pi s^2_W}U_W(q^2)\nonumber\\
&&+{\alpha c^2_W\over\pi s^2_W}ln[{1-cos\theta\over2}]
ln{q^2\over M^2_W}-{\alpha v^2_e\over4\pi s^2_Wc^2_W}
ln[{1+cos\theta\over1-cos\theta}]
ln{q^2\over M^2_Z}
\label{Rl}\eqa

\bqa
V^{S}_{\gamma Z,l\mu}(q^2,\theta)&=&V^{S}_{Z\gamma ,l\mu}(q^2,\theta)
={\alpha\over8\pi s_W c_W}U_W(q^2)
-[{\alpha v_e(1-v^2_e)\over128\pi s^3_Wc^3_W}
+{\alpha v_e\over8\pi s_Wc_W}]U_Z(q^2)\nonumber\\
&&
-{\alpha c_W\over2\pi s_W}U_W(q^2)
+{\alpha c_W\over\pi s_W}ln[{1-cos\theta\over2}]
ln{q^2\over M^2_W}
-{\alpha v_e(1-v_e^2)\over32\pi s^3_Wc^3_W}ln[{1+cos\theta
\over1-cos\theta}]
ln{q^2\over M^2_W}
\label{Vl}
\eqa

\bqa
\tilde{\Delta}^{S}_{\alpha,lu}(q^2,\theta)&=&
-~{\alpha\over12\pi}U_W(q^2)
+{\alpha(2-v^2_e-v^2_u)\over64\pi s^2_Wc^2_W}U_Z(q^2)
+{\alpha\over2\pi}U_W(q^2)\nonumber\\
&&-{\alpha\over\pi}ln[{1+cos\theta\over2}]
ln{q^2\over M^2_W}-{3\alpha(1-v^2_e)
(1-v^2_u)\over512\pi s^4_Wc^4_W}
ln[{1+cos\theta\over1-cos\theta}]
ln{q^2\over M^2_Z}\nonumber\\
&&
-\frac{\alpha}{24\pi s_W^2} \ln {q^2\over m^2_t}
\left[
(3-2s^2_W)
\frac{m_t^2}{M_W^2}+2s_W^2\frac{m_b^2}{M_W^2}
\right]
\label{DAu}
\eqa

\bqa
R^{S}_{lu}(q^2,\theta)&=&{\alpha\over4\pi s^2_W}
(1-{s^2_W\over3})U_W(q^2)
-{\alpha(2+3v^2_e+3v^2_u)\over64\pi s^2_Wc^2_W}U_Z(q^2)
-{\alpha c^2_W\over2\pi s^2_W}U_W(q^2)\nonumber\\
&&+{\alpha c^2_W\over\pi s^2_W}ln[{1+cos\theta\over2}]
ln{q^2\over M^2_W}+{\alpha v_e v_u\over4\pi s^2_Wc^2_W}
ln[{1+cos\theta\over1-cos\theta}]
ln{q^2\over M^2_Z}\nonumber\\
&&
+\frac{\alpha}{16\pi s^2_W} \ln {q^2\over m^2_t}
\left[
 \left(1+\frac{4s_W^2}{3}\right) \frac{m_t^2}{M_W^2}
+\left(1-\frac{4s^2_W}{3}\right) \frac{m^2_b}{M^2_W}
\right]
\label{Ru}\eqa

\bqa
V^{S}_{\gamma Z,lu}(q^2,\theta)&=&{\alpha(3+2c^2_W)
\over24\pi s_W c_W}U_W(q^2)
-[{\alpha v_e(1-v^2_e)\over128\pi s^3_Wc^3_W}
+{\alpha v_u\over12\pi s_Wc_W}]U_Z(q^2)
-{\alpha c_W\over2\pi s_W}U_W(q^2)\nonumber\\
&&+{\alpha c_W\over\pi s_W}ln[{1+cos\theta\over2}]
ln{q^2\over M^2_W}
+{\alpha v_u(1-v_e^2)\over32\pi s^3_Wc^3_W}ln[{1+cos\theta
\over1-cos\theta}]
ln{q^2\over M^2_W}\nonumber\\
&&-{\alpha c_W\over12 \pi s_W} \ln {q^2\over m^2_t}
\left(
\frac{m^2_t}{M^2_W}-\frac{m_b^2}{M_W^2}
\right)
\label{Vgzu}
\eqa

\bqa
V^{S}_{Z\gamma,lu}(q^2,\theta)&=&{\alpha(3-2s^2_W)
\over24\pi s_W c_W}U_W(q^2)
-[{3\alpha v_u(1-v^2_u)\over256\pi s^3_Wc^3_W}
+{\alpha v_e\over8\pi s_Wc_W}]U_Z(q^2)
-{\alpha c_W\over2\pi s_W}U_W(q^2)\nonumber\\
&&+{\alpha c_W\over\pi s_W}ln[{1+cos\theta\over2}]
ln{q^2\over M^2_W}
+{3\alpha v_e(1-v_u^2)\over64\pi s^3_Wc^3_W}ln[{1+cos\theta
\over1-cos\theta}]
ln{q^2\over M^2_W}\nonumber\\
&&
-{\alpha\over 16\pi s_W c_W} \ln {q^2\over m^2_t}
\left(1-\frac{4s^2_W}{3}\right)(\frac{m^2_t}{M^2_W}
-\frac{m_b^2}{M_W^2})
\label{Vzgu}
\eqa

\bqa
\tilde{\Delta}^{S}_{\alpha,ld}(q^2,\theta)&=&
-~{\alpha\over6\pi}U_W(q^2)+
{\alpha(2-v^2_e-v^2_d)\over64\pi s^2_Wc^2_W}U_Z(q^2)
+{\alpha\over2\pi}U_W(q^2)\nonumber\\
&&-{\alpha\over\pi}ln[{1-cos\theta\over2}]
ln{q^2\over M^2_W}+{3\alpha(1-v^2_e)
(1-v^2_d)\over256\pi
s^4_Wc^4_W}ln[{1+cos\theta\over1-cos\theta}]
ln{q^2\over M^2_Z}\nonumber\\
&&-{\alpha\over24\pi s^2_W}
(ln{q^2\over m^2_t})
~[~s^2_W({m^2_t\over M^2_W})
+(3-s^2_W)({m^2_b\over M^2_W})~]
\label{DAd}
\eqa

\bqa
R^{S}_{ld}(q^2,\theta)&=&{\alpha\over4\pi s^2_W}
(1-{2s^2_W\over3})U_W(q^2)
-{\alpha(2+3v^2_e+3v^2_d)\over64\pi s^2_Wc^2_W}U_Z(q^2)
-{\alpha c^2_W\over2\pi s^2_W}U_W(q^2)\nonumber\\
&&+{\alpha c^2_W\over\pi s^2_W}ln[{1-cos\theta\over2}]
ln{q^2\over M^2_W}-{\alpha v_ev_d\over4\pi s^2_Wc^2_W}
ln[{1+cos\theta\over1-cos\theta}]
ln{q^2\over M^2_Z}\nonumber\\
&&+{\alpha\over16\pi s^2_W}(ln{q^2\over m^2_t})
~[(1-{2s^2_W\over3})({m^2_t\over M^2_W})
+(1+{2s^2_W\over3})({m^2_b\over M^2_W})~]
\label{Rd}
\eqa

\bqa
V^{S}_{\gamma Z,ld}(q^2,\theta)&=&{\alpha(3+4c^2_W)
\over24\pi s_W c_W}U_W(q^2)
-[{\alpha v_e(1-v^2_e)\over128\pi s^3_Wc^3_W}
+{\alpha v_d\over24\pi s_Wc_W}]U_Z(q^2)
-{\alpha c_W\over2\pi s_W}U_W(q^2)\nonumber\\
&&+{\alpha c_W\over\pi s_W}ln[{1-cos\theta\over2}]
ln{q^2\over M^2_W}
-{\alpha v_u(1-v_e^2)\over32\pi s^3_Wc^3_W}ln[{1+cos\theta
\over1-cos\theta}]
ln{q^2\over M^2_W}\nonumber\\
&&+{\alpha c_W\over24\pi s_W}(ln{q^2\over m^2_t})
~[~({m^2_t\over M^2_W})
-({m^2_b\over M^2_W})~]
\label{Vgzd}
\eqa

\bqa
V^{S}_{Z\gamma,ld}(q^2,\theta)&=&{\alpha(3-4s^2_W)
\over24\pi s_W c_W}U_W(q^2)
-[{3\alpha v_d(1-v^2_d)\over128\pi s^3_Wc^3_W}
+{\alpha v_e\over8\pi s_Wc_W}]U_Z(q^2)
-{\alpha c_W\over2\pi s_W}U_W(q^2)\nonumber\\
&&+{\alpha c_W\over\pi s_W}ln[{1-cos\theta\over2}]
ln{q^2\over M^2_W}
-{3\alpha v_e(1-v_d^2)\over32\pi s^3_Wc^3_W}ln[{1+cos\theta
\over1-cos\theta}]
ln{q^2\over M^2_W}\nonumber\\
&&
-{\alpha\over16\pi s_W c_W}(ln{q^2\over m^2_t})(1-{2s^2_W\over3})
[({m^2_t\over M^2_W})
-({m^2_b\over M^2_W})~]
\label{Vzgd}
\eqa

\noindent
where $v_f\equiv1-4s^2_W|Q_f|$.

Before moving to a detailed numerical investigation of the Sudakov
effect on the various experimental observables of the process, there
are a few preliminary general remarks that, we feel, might be relevant.
In particular, the following points should be mentioned:\par
I) The universal and the non universal sets, that we have grouped in
the various equations, should be in our opinion, \underline{separately}
gauge-independent. Given the fact
that for all the \underline{overall} listed form factors,
by construction, this property
holds true, the same feature must obtain both for the
\underline{overall} contributions of  \underline{non universal}
kind and for the \underline{overall} contributions of  
\underline{universal} kind that are considered. 
Gauge dependence can only affect,
separately, the \underline{universal} contributions coming from the
$2W$ vertex and from the $W$ box. But their special combination, 
that builds the same universal contribution $U_W(q^2)$ produced by the
(\underline{gauge-independent}) single $W$ vertex, must necessarily be
gauge-independent as well. This fact
reproduces an analogous well known property\cite{17} of the $2W$
vertex. In fact, the so-called "pinch" component\cite{18} of this
vertex is gauge-dependent, and combines with a corresponding
gauge-dependent part in the various ($\gamma,Z$) self-energies to
make up gauge-independent quantities that Degrassi and Sirlin call
"gauge-independent" self-energies, that produce the correct 
asymptotic RG logarithmic behaviour of the running couplings. 
In our case, a combination of the $2W$ vertex with a \underline{box}
generates the "correct" gauge-independent asymptotic Sudakov logarithmic
behaviour (that exponentiates).\par
II) From the practical point of view of the validity of a perturbative
expansion truncated at one loop, we believe that one should consider
the terms in the various brackets and discuss the various effects
separately in all observables. For instance, a cancellation might arise
between the $\theta$-independent and the $\theta$-dependent
contributions if they were both large and of opposite sign. This, we
believe, would not make a one loop approximation reliable.\par
III) As a rather academic feature, we believe that it should be
stressed that, for all "light" (massless) fermion production processes,
there exists a "Magic Energy" where the $\theta$-independent functions 
$U_W(q^2)$, $U_Z(q^2)$ both vanish. This corresponds to the choice

\bq
ln{q^2\over M^2_W}\simeq ln{q^2\over M^2_Z} =3
\label{MB}\eq
\noindent
that selects the "Magic Energy" 

\bq
\sqrt{q^2}\simeq 360 ~{\rm GeV}
\label{MBE}\eq

Clearly, in the vicinity of this energy, \underline{all} the
logarithmic Sudakov contribution for massless fermions is produced by
the $\theta$-dependent, non universal components of the weak boxes
(which reduces essentially, from the numerical point of view, to the
contribution from the $W$ box). For bottom production at this energy,
an extra amount of Yukawa Sudakov logarithms must be added (for top
production, we believe that $360~{\rm GeV}$ is definitely not an "asymptotic"
energy, and the validity of an asymptotic expansion is strongly
debatable; we shall only treat top production in this paper in the CLIC
$\sqrt{q^2}=3~{\rm TeV}$ regime).\par
After these general remarks, we are now ready to perform a numerical
investigation of the various asymptotic Sudakov logarithms effects on
all the observables of the process. This will be done in the next
Section 3.

\section{Effects of the different Sudakov logarithms on the
experimental observables}

Having examined the way in which the $\theta$-independent Sudakov
logarithms always group, for final massless fermions, in the
combination ($3ln q^2 -ln^2q^2$), for which precise rules
exist\cite{Melles} that make its resummation known (these are also
available for the massive Yukawa contributions), we shall now proceed
to the calculation of the Sudakov effects on various observables at one
loop. With this aim, we shall consider the effects as due to three
separate categories of terms : those which arise from
$\theta$-independent quantities and enter in the two possible
combinations $U_W(q^2)$, $U_Z(q^2)$ eqs.(\ref{UW},\ref{UZ}); those which
arise from $\theta$-dependent terms (denoted "$\theta$ S"), 
and those which are of massive
Yukawa origin (denoted "YU"). 
Since we are only interested in these specific
contributions, we shall only write their (relative) effects on the
different cross sections and their (absolute) effects on the different
asymmetries, denoting with a "NS" ($\equiv$ Non Sudakov) symbol all the
remaining part of the various observables. For sake of comparison, we
shall also include in our formulae the relative and absolute effects
due to the linear logarithms of Renormalization Group (RG) origin,
already computed (e.g. \cite{log}). Also, for simplicity, we shall
group together the two $W$ and $Z$ functions writing $U_W\simeq
U_Z\simeq U$ ($M_W\simeq M_Z \equiv M$), 
which creates a small numerical difference that will be
irrelevant for the specific purposes of this paper, given the fact that
the $Z$ term is much smaller than the $W$ one. We shall consider as
realistic observable final states those which contain a $\mu^+\mu^-$
(or , also, a $\tau^+\tau^-$ pair), a $b\bar b$ pair, a $t\bar t$ pair.
Also, the cross section for production of the five "light" quarks
$\sigma_5$ will be considered. Note that for what concerns top
production our formalism, strictly speaking, only applies to energies
in the CLIC range\cite{mt}, and for this reason this process will not be
studied in the LC regime. We shall restrict our attention on cross
sections for production of a single final state $f\bar f$ ($\sigma_f$),
on $\sigma_5$,  and on forward-backward asymmetries ($A_{FB,f}$) 
and also longitudinal
polarization asymmetries ($A_{LR,f}$) whose conventional definitions 
are recalled in Appendix B.
Starting from the expressions given in Appendix B and from
Eqs.(\ref{DAl}-\ref{Vzgd}) 
it is a relatively straightforward task to derive the
various Sudakov effects. We shall write them in what follows replacing
the theoretical input weak parameters by their experimental values, to
make the different numerical size of the various terms immediately
evident.\par
We now list the final expressions for the various observables. They
read:

\bqa
\sigma_{\mu}&=&\sigma^{NS}_{\mu}\{1+{\alpha(M)\over\pi}[
(0.645~ln{q^2\over M^2})_{RG}+(1.51~U(q^2))
+(5.49~ln{q^2\over M^2})_{\theta S}]
\eqa

\bqa
A_{FB,\mu}&=&A^{NS}_{FB,\mu}+{\alpha(M)\over\pi}[
(-1.07~ln{q^2\over M^2})_{RG}+(0.021~U(q^2))
+(2.80~ln{q^2\over M^2})_{\theta S}]
\eqa

\bqa
A_{LR,\mu}&=&A^{NS}_{LR,\mu}+{\alpha(M)\over\pi}[
(-3.58~ln{q^2\over M^2})_{RG}+(0.92~U(q^2))
+(5.13~ln{q^2\over M^2})_{\theta S}]
\eqa

\bqa
\sigma_{b}&=&\sigma^{NS}_{b}\{1+{\alpha(M)\over\pi}[
(-5.30~ln{q^2\over M^2})_{RG}+(2.39~U(q^2))\nonumber\\
&&+(15.01~ln{q^2\over M^2})_{\theta S}-(2.10~ln{q^2\over m^2_t})_{YU}]
\eqa

\bqa
A_{FB,b}&=&A^{NS}_{FB,b}+{\alpha(M)\over\pi}[
(-1.11~ln{q^2\over M^2})_{RG}+(0.098~U(q^2))\nonumber\\
&&
+(4.25~ln{q^2\over M^2})_{\theta S}-(0.09~ln{q^2\over m^2_t})_{YU}]
\eqa

\bqa
A_{LR,b}&=&A^{NS}_{LR,b}+{\alpha(M)\over\pi}[
(-3.71~ln{q^2\over M^2})_{RG}+(0.72U~(q^2))\nonumber\\
&&
+(5.30~ln{q^2\over M^2})_{\theta S}-(0.60~ln{q^2\over m^2_t})_{YU}]
\eqa

\bqa
\sigma_{5}&=&\sigma^{NS}_{5}\{1+{\alpha(M)\over\pi}[
(-3.26~ln{q^2\over M^2})_{RG}+(2.08~U(q^2))\nonumber\\
&&
+(7.22~ln{q^2\over M^2})_{\theta S}-(0.30~ln{q^2\over m^2_t})_{YU}]
\eqa

\bqa
R_{b}\equiv{\sigma_{b}\over\sigma_{5}}
&=& R_{b}^{NS} +{\alpha(M)\over\pi}[
(-0.29~ln{q^2\over M^2})_{RG}+(0.045~U(q^2))\nonumber\\
&&
+(1.12~ln{q^2\over M^2})_{\theta S}-(0.26~ln{q^2\over m^2_t})_{YU}]
\eqa

\bqa
A_{LR,5}&=&A^{NS}_{LR,5}+{\alpha(M)\over\pi}[
(-4.16~ln{q^2\over M^2})_{RG}+(0.92~U(q^2))\nonumber\\
&&
+(3.67~ln{q^2\over M^2})_{\theta S}-(0.13~ln{q^2\over m^2_t})_{YU}]
\eqa

\bqa
\sigma_{t}&=&\sigma^{NS}_{t}\{1+{\alpha(M)\over\pi}[
(-1.64~ln{q^2\over M^2})_{RG}+(1.80~U(q^2))\nonumber\\
&&
+(1.18~ln{q^2\over M^2})_{\theta S}-(3.55~ln{q^2\over m^2_t})_{YU}]
\eqa

\bqa
A_{FB,t}&=&A^{NS}_{FB,t}+{\alpha(M)\over\pi}[
(-0.88~ln{q^2\over M^2})_{RG}+(0.06~U(q^2))\nonumber\\
&&
-(0.93~ln{q^2\over M^2})_{\theta S}+(0.15~ln{q^2\over m^2_t})_{YU}]
\eqa

\bqa
A_{LR,t}&=&A^{NS}_{LR,t}+{\alpha(M)\over\pi}[
(-4.06~ln{q^2\over M^2})_{RG}+(1.01~U(q^2))\nonumber\\
&&
+(0.76~ln{q^2\over M^2})_{\theta S}+(0.95~ln{q^2\over m^2_t})_{YU}]
\eqa

In the case of top production, other observables can be added that
depend on the final top helicity. They are listed in the second of
Ref.\cite{mt}. As it was already remarked in that reference, these
extra observables only differ from their corresponding "unpolarized
top" quantities by linear Sudakov logarithms of $\theta$-dependent box
origin. Therefore the conclusions concerning the $\theta$-independent
terms will remain unchanged. A more complete discussion about the
$\theta$-dependent effects could be given, but it seems to us beyond
the specific purposes of this paper. We shall 
defer to a dedicated forthcoming paper devoted to top
production \cite{top-2} for more details.\par
We may try to compare the above expressions with results obtained
by other authors e.g \cite{DenPo,KPS,Melles}. However the
comparison is not obvious because the photonic part is treated
differently in these papers and the observables are often not
defined in the same way. One can nevertheless identify the main terms.
The comparison is easier with the results in
ref\cite{KPS}. They differ only by the inclusion
of the small $\theta$-dependent contribution 
from the $\gamma Z$ boxes and are indeed very close to our
results, as one can easily verify.\par
We are now ready for a detailed numerical investigation of the Sudakov
effects on the listed observables at one loop. With this aim, we divide
our analysis into two parts, separately devoted to the two cases of
energies in the LC($\sqrt{q^2}\simeq500~{\rm GeV}$) and in the 
CLIC($\sqrt{q^2}\simeq3~{\rm TeV}$) regime. Our main conclusions can be
summarized as follows:\\

I) \underline{LC regime ($\sqrt{q^2}\simeq500~{\rm GeV}$)}\\

At $\sqrt{q^2}\simeq500~{\rm GeV}$, the Sudakov effects act on the various
observables in quite a different way. We have made the following general
classification:\\

a) \underline{Cross sections}. 
The relative effect in permille on the muon cross
section is, to good approximation, a negative five from the
$\theta$-independent and a positive forty-six 
from the $\theta$-dependent
term. For the "light" quark cross section $\sigma_5$ it is a negative
seven ($\theta$-independent) and a positive sixty-one
($\theta$-dependent). For bottom production, the relative effect on the
cross section is a negative eight ($\theta$-independent) and a
positive hundred-twentysix ($\theta$-dependent). One has also, in this
case, a negative relative effect of eleven permille coming from the
extra linear Sudakov logarithm of Yukawa origin.\par
The general comment that can be made at this point is that, for all the
considered cross sections, the effect of the non universal,
$\theta$-dependent subleading Sudakov logarithm is, at one loop, by far
larger than that of the $\theta$-independent combinations 
$U_W(q^2),~U_Z(q^2)$, and systematically of opposite sign. This is,
somehow, unfortunate since a resummation prescription for the 
$\theta$-dependent logarithms does not seem to exist at the moment
\cite{Melles}. In the LC range, this might not represent a problem for
a one loop approximation if one considers the relative one percent as a
reasonable experimental achievement for $\sigma_{\mu}$ and $\sigma_5$.
In this case, relative effects at one loop around five
percent might be tolerated, with some warning in the case of $\sigma_5$.
For bottom production, if one expect an experimental accuracy of a few
percent, a thirteen percent effect would still be acceptable. If the
experimental precision were higher than the previous qualitative
estimates given here, the necessity of a two loop calculation for the 
$\theta$-dependent contribution would become imperative. Note,
accidentally, that the effects of the resummable terms $U_W(q^2),~U_Z(q^2)$,
are in this case extremely small at the one loop level, so that in
their case a one loop approximation seem to us completely reliable.\\

b) \underline{Longitudinal polarization asymmetries}. This case
presents strong similarities with that of the corresponding cross
sections, and therefore we treat it immediately in succession. The
\underline{absolute} effect (in permille) on the muon asymmetry,
$A_{LR,\mu}$ is a negative three ($\theta$-independent) and a positive
forty-three ($\theta$-dependent). 
For the five light quarks case $A_{LR,5}$, the
absolute effect is a negative three ($\theta$-independent) and a
positive thirty-one ($\theta$-dependent). For bottom production, 
$A_{LR,b}$, the two effects are respectively minus two and plus
forty-five. Again, one notices a strong dominance at one loop of the
positive $\theta$-dependent terms with respect to the negative 
$\theta$-independent ones, just as in the case of cross sections. The
reliability of the one loop approximation will strongly depend on the
aimed experimental accuracies of the measurements. If these will remain
at the (absolute) few percent level, there should be no problem for the
approximation, while higher experimental accuracies would make a
two-loop calculation of the $\theta$-dependent terms highly
"desirable".\\

c)\underline{Forward-backward asymmetries}. These specific observables
present a peculiar feature that extremizes the previously remarked 
"$\theta$-dependent logarithms dominance". In fact, in their case,
an accurate numerical calculation shows that, \underline{independently
of the considered final state}, the coefficient of the 
$\theta$-independent $\simeq U_W(q^2),~U_Z(q^2)$ terms \underline{is
always, essentially, negligibly 
small}, i.e. much smaller than that of the $\theta$-dependent term 
and well below the absolute percent level.
In Appendix A we have tried to derive in some detail this apparently
non trivial fact, which seems to arise from the multiplets assignment
of the fermions in $SU(2)\times U(1)$. Numerically and to a good
approximation this absolute effect is (in permille) always negative and
in magnitude well less than one for final muons and one for $b's$. 
\underline{This
feature will persist at the higher energies involved at CLIC}, where it
will apply also to top production (that we do not treat at LC
energies), \underline{and seems to be a very general property of this type of
observables}. The consequence is that, at asymptotic energies, the only
Sudakov logarithms that must be retained at one loop in the
forward-backward asymmetries are the 
$\theta$-dependent ones of box origin. This generates a strange
situation of "box dominance" for what concerns this type of virtual
effects, totally opposite to the situation e.g. met on top of the $Z$
resonance.\footnote{As we already anticipated in
Eqs.(\ref{MB},\ref{MBE}), 
a similar and rather peculiar
feature of the Sudakov logarithms arise at
$\sqrt{q^2}\simeq360~{\rm GeV}\simeq4~M_Z$. Here the $\theta$-independent
term vanishes \underline{exactly} at one loop, so that the full effect
is produced, at that energy, by weak boxes. This situation is again
just opposite to that met at the $Z$ peak where boxes could be safely
ignored, and affects not only the forward-backward asymmetries but
\underline{all} observables at this special energy value.}\par
For what concerns the validity of a one-loop approximation, the 
$\theta$-dependent absolute effects are always positive and equal, in
permille, to twenty-four (final muons) and thirty-six (final bottom). At
the percent level of experimental accuracies, this seems to us to make
the approximation reliable, even in cases of reasonable
improvements in the experimental precision.\\

II) \underline{CLIC regime ($\sqrt{q^2}=3~{\rm TeV}$)}\\

For energies of about $3~{\rm TeV}$, as those aimed for in
the first phase of the
future CERN CLIC Collider, we have repeated the previous analysis
including also top production for which CLIC energies can be safely
considered as "asymptotic". We list here the results that we have
found, quoting the $\theta$-independent term first, then giving the 
$\theta$-dependent term effect and finally, whenever involved, giving
the Yukawa term contribution (in percent this time, relative for cross
sections and absolute for asymmetries).\\

a) \underline{Cross sections}. 
For final muons, we get a negative ten and a
positive nine (percent). For final light quarks ($\sigma_5$), there
appear a negative fourteen and a positive twelve. For bottom
production, a negative sixteen and a positive twentysix (plus a
negative three of Yukawa origin). For top production, a negative
twelve and a positive two (with a negative five of Yukawa origin).\par
As one sees, the situation at CLIC is strongly different from the
corresponding one at LC. The role of the $\theta$-independent
terms is now slightly more relevant than that of the $\theta$-dependent
ones for both muon and for light quarks production and largely
dominant for top production, it remains less relevant only for bottom
production.\\

For what concerns the validity of a one-loop approximation, the
situation seems to us to be, in a certain sense, 
embarassing, and also final
state dependent. For muon and light quark production, one might take
the pragmatic attitude of considering the \underline{overall} Sudakov
effect, obtained by summing the negative $\theta$-independent and the
positive $\theta$-dependent one. This sum is actually small (a few
percent) and apparently under control. A more cautious point of view
though, that we personally share, is that one is dealing here with
\underline{two} large and opposite effects, that are both separately
gauge-independent and of rather different origin, the positive one
being completely non-universal and $\theta$-dependent. We do not see
any obvious reason why the two large and independent effects should
still cancel e.g. at the next two-loop level. Thus, 
in our opinion, one should
compute them \underline{both} to higher order.
This would not represent
a problem for the $\theta$-independent contribution, for which
resummation prescriptions exist\cite{Melles}. But, as we already said,
these prescriptions are unclear for the $\theta$-dependent term. Given
its rather large size, a calculation of this quantity at the next
two-loop level seems to us, least to say, unavoidable.\par
Note that, in our opinion, until a two-loop calculation of the latter
term has been performed, resumming the $\theta$-independent effect
only, leaving the other terms at the one loop level, could
worsen the situation. This procedure might in fact reduce the resummed
negative contribution, leaving a much larger positive dominant term.
Unfortunately, it seems to us that for light fermion production cross
sections at CLIC energies, a resummation of the pure
$\theta$-independent terms, although theoretically valid and remarkable,
does not provide the full answer to the need of a reliable, complete
theoretical prediction.\par
This conclusion remains unchanged, in our opinion, also for bottom
production, from inspection of the numerical effects that we have
shown. The only apparent evasion of this negative statement is provided
by the cross section for top production. Here the $\theta$-independent
effect dominates, while the other one is small and limited (two
percent). The reason for the weakness of the $\theta$-dependent term 
is the fact that its leading contribution, the $W$ box diagram for
$t\bar t$ production, has an angular distribution 
$\simeq ln[(1+cos\theta)/2]$ which is peaked backward (as one can guess
from the diagram (2) of Fig.1d) and interfers very little with the
forward peaked Born term; in the case of $b\bar b$ production the
diagram (1) of Fig.1d, on the contrary, is peaked forward and
interfers strongly with the Born term.
For $t\bar t$ production in this situation, 
one could safely approximate the cross section with the
one-loop calculation, resumming only the $\theta$-independent term. This
process could therefore be already satisfactorily calculated, without
extra theoretical efforts, by a suitable combination of different
existing formulae for the $\theta$-independent\cite{Melles} and the
$\theta$-dependent \cite{mt} contributions.\\

b) \underline{Polarization asymmetries}.\\
One finds again, as in the LC case, a situation that is similar to that
of the corresponding cross sections. The absolute numbers (in percent)
are for final muons, $A_{LR,\mu}$, minus six and plus nine; for light
quarks, $A_{LR,5}$, minus six and plus six; for final bottom, minus
five and plus nine; for final top, minus seven and plus one. Assuming
a (approximately) percent level for the related experimental precisions, we
believe that the same conclusions, that were just drawn 
in (IIa) for what
concerns the one-loop approximation at CLIC energies for the various
light fermions, bottom, top cross sections, still apply for all the
corresponding longitudinal polarization asymmetries.\\

c) \underline{Forward-backward asymmetries}.\\

At CLIC energies, the absolute overall contribution of the
$\theta$-independent Sudakov terms is, for both massless and massive
fermion production, systematically irrelevant
(one permille for muons, few permille
for either light or massive quarks) at the level of realistic
expectable experimental accuracy. This is in agreement with our
general previous observation, that will be discussed separately in
Appendix A. The box $\theta$-dependent absolute effects are,
respectively, five percent (final muons), seven percent (bottom
production), minus two percent (top production). With an experimental
accuracy of one percent for muons and top, and of a few percent for
bottom production, a one-loop approximation seems to us fully
acceptable. In this case, the relevant effect would be fully provided
by existing one-loop calculations \cite{log,mt} 
of the angular dependent
component of the terms, without need of any extra theoretical
effort.\par
We have thus completed our numerical analysis in the two (LC, CLIC)
different considered "asymptotic" energies. The main results and
conclusions are summarized in the forthcoming and final Section 4.

\section{Conclusions}

We have performed in this paper a systematic analysis of the weak
Sudakov logarithmic effects at one loop in the 't Hooft $\xi=1$ gauge
for a large class of experimental observables, in two different energy
configurations that correspond to the regimes to be explored at the
next linear colliders LC ($500~{\rm GeV}$) and CLIC ($3~{\rm TeV}$). 
We have
divided the set of effects into two essentially different
gauge-independent subsets. The first one is "angular independent", is
universal and 
comes from vertices and boxes; the second
one is ``angular dependent", is not universal and only comes
from boxes. The main motivation of our analysis was that of studying
the specific effects of this last term, for which no clean resummation
prescription beyond the one loop level seems to exist at the
moment\cite{Melles}.\par
Our numerical analysis has been explicitely performed at two selected
energies, $500~{\rm GeV}$ and $3~{\rm TeV}$, but it can be 
easily repeated for any
arbitrary value e.g. beyond $500~{\rm GeV}$ or $3~{\rm TeV}$. 
This has been
summarized pictorially in Figs.(2-11), where we have shown the various
contributions from the angular independent and from the angular
dependent term on all the chosen observables, when the energy varies.
For sake of completeness we have also enclosed the universal linear
logarithmic contribution of RG origin, which has been computed in
previous references \cite{log,mt} 
and does not seem to become "dangerous"
at the considered energies. From inspection of these Figures, one sees
e.g. that our conclusions remain essentially valid when $\sqrt{q^2}$
ranges between $\simeq3$ and $\simeq5$ TeV, a possible larger CLIC
range, or between $500~{\rm GeV}$ and $1~{\rm TeV}$ in the LC case.\par
Our general conclusion is that, for each considered observable, in both
the considered energy configurations, the role of the angular dependent
term is \underline{always} essential. In the cases of cross sections
and longitudinal polarization asymmetries, our analysis has led to very
similar conclusions for the "corresponding" quantities (i.e. the cross
section for production of a certain final state and the related
longitudinal polarization asymmetry). In practice, at LC, the angular
dependent logarithms are dominant but "small" i.e. "under control" at
one-loop, assuming an experimental accuracy of the percent size. At
CLIC, a strong cancellation appears at one loop between the "large"
(i.e. $\gtrsim$ 10 \%) negative 
angular independent effect and the "large"
positive angular dependent one. So both should not be taken,
in our opinion, in the
one-loop approximation, which does not represent a problem for the
first term, but requires a new calculation at least at two-loops for
the second one. An exception to this statement is provided by the cross
section for top production, the only case that we found where the
angular dependent effect turns out to be negligible.\par
A completely separate role is played by the forward-backward
asymmetries. In these observables, independently of the considered
"high" energy, the angular independent effect at one-loop is
essentially vanishing, for reasons that seem to be accidental.
Thus the angular dependent Sudakov logarithm remains, for these special
quantities, the only relevant effect. Luckily, if one assumes realistic
experimental accuracies, this effect appears to be under control both
at LC and at CLIC energies, which would allow to avoid a hard two-loop
calculation in all cases.\par
As a matter of fact, the need of a two-loop calculation of the angular
dependent term is only appearing for calculations of cross sections (of
which longitudinal polarization asymmetries are essentially a special
case). The possibility that a simpler calculational approach can be
found for these well defined cases is, at the moment, being
investigated.\par
A numerical simplification has occured, in fact, in our approach and we
want to discuss it now briefly. In the calculation of the size of the
effect of the $\theta$-independent Sudakov terms on the various
observables we have fully retained the asymptotic expressions given in
Eqs.(\ref{DAl})-(\ref{Vzgd}). In the two limiting situations of forward
and backward scattering, $\cos\theta\to\pm1$, the asymptotic
expressions formally diverge like a logarithm. Clearly, this would not
be the case if we had used the complete expression, which would be
necessary in the low $q^2$ range. For the specific purposes of this
paper, where only $\theta$-integrated quantities have been considered
and estimates of effects are essentially indicative (i.e. aiming at
identifying "dangerous" potential contributions), this approximation
seems to us satisfactory in the large $q^2$ regime in which we are
interested. First of all, a logarithmic singularity produces in any
case a finite integrated quantity. In second place, one must remember
that in a realistic experiment there is always a finite value of the
scattering angle, $\theta=\theta_0$, below which no experimental
observation is allowed. Starting from these considerations we have
first recomputed the integration of the $\theta$-dependent logarithms
with a cut at $\pm\cos\theta_0$, and compared these values with those
obtained performing the full integration that would correspond to
$\theta_0=0$. More precisely, we have considered the two quantities 
which appear in the $W$ box contribution to the cross sections:
%


\bq
I_1=\int^{\cos\theta_0}_{-\cos\theta_0} 
d\cos\theta ~(1+\cos^2\theta)ln(1\mp\cos\theta)
\label{I1}\eq
\bq
I_2=\mp\int^{\cos\theta_0}_{-\cos\theta_0} 
d\cos\theta ~\cos\theta~ln(1\mp\cos\theta)
\label{I2}\eq
\noindent
These can be computed analytically leading to the expressions

\bqa
I_1&=&-\frac{8}{3}\cos\theta_0 -\frac{2}{9} \cos^3\theta_0 +
\frac{1}{3}(-4+3\cos\theta_0+\cos^3\theta_0)\ln(1-\cos\theta_0)+ \nonumber \\
&+& \frac{1}{3}(4+3\cos\theta_0+\cos^3\theta_0)\ln(1+\cos\theta_0)
\label{I1a}
\eqa
\bq
I_2=\cos\theta_0-\frac{1}{2}\sin^2\theta_0\ln\frac{1+\cos\theta_0}{1-\cos\theta_0}
\label{I2a}
\eq
\noindent
and for example, one finds 
$I_1=~-1.04,~ -0.90,~-0.67,~-0.44$, and $I_2=~1,~ 0.91,~ 0.74,~ 0.54$
when $\theta_0=~0,~10^0,~20^0,~30^0$,
respectively. As one sees, the "cut" quantities, for 
values of $\theta_0$ as large as $\simeq20^0$, only differ
from the complete integration by a relative $20-30$\%
difference, and will   
\underline{essentially} reproduce its main features, so that this cut
effect will be irrelevant for our conclusions. In
this region, our "logarithmic approximation" should be satisfactory. In
fact, in the expressions to be integrated we made the assumption

\bq
\theta>>{M_W\over\sqrt{q^2}}
\label{mass}\eq	
\noindent
which ensures that the terms $ln^2(t/M^2_W),~ln^2(u/M^2_W)$
are large and can safely be estimated by neglecting their
$q^2$-independent parts.\par
At CLIC energies, the r.h.s. of Eq.(\ref{mass}) is equal to $\simeq1.5$
degree while in the LC range it reaches a value of about $9$ degrees
(or less, for $\sqrt{q^2}>500$ GeV). In both cases, from what
previously shown, an estimate of the angular cut that must be performed
in the different observables, based on the logarithmic approximation 
truncated, say, at a corresponding
realistic cut, would reproduce "essentially" the numbers that we gave,
to a degree of accuracy that should be fixed by a dedicated analysis of
the special experimental features of the related experiments.\par
We notice at this point that the same logarithmic approximation that we
followed was also used in Ref.\cite{KPS}, where a detailed numerical
analysis has been performed, with precise numbers, for some of 
the observables that we considered. 
Since the analysis of Ref.\cite{KPS} also includes,
as we already mentioned, the $\theta$-dependent contribution from the
$\gamma Z$ boxes, slight differences appear in the
various results.\par
A final comment should now be made concerning the role played by other
possible asymptotic logarithms in the examined processes. The (linear
ones) of RG origin have been listed in our formulae, and the reader can
very easily verify that their effect at one loop will never be
"dangerous" at the considered energies. There is another interesting
possibility, due to virtual Sudakov effects at one loop of
supersymmetric origin. They have been exhaustively discussed for the
MSSM case, in the case $f\neq t$ 
in the few ${\rm TeV}$ regime in a recent paper\cite{slog}. They
turn out to be only linearly logarithmic, and systematically "under
control" at CLIC energies with a possible exception for bottom
production (assuming a typical SUSY mass of few hundred
${\rm GeV}$). So, for $f\neq t$ production, 
SUSY does not add technical
problems at the theoretical one-loop level in the MSSM case. Note that,
also in the MSSM case, the extra SUSY $\theta$-independent contributions
to the considered forward-backward asymmetries are systematically
negligible, exactly as in the SM case.\par
In the case $f=t$, discussed in the second of Ref.\cite{mt}, the
situation is slightly less straightforward. The size of the linear
Sudakov logarithm, that contains a large component of Yukawa origin,
depends strongly on the SUSY parameter $\tan\beta$. For the lowest allowed
values of $\tan\beta$, its numerical value in the cross section can be larger
than the ten percent "safety" limit. A strong reduction of the effect
is, though, achievable by adding in the asymptotic expansion a
reasonable extra SUSY constant~\footnote{
As a matter of fact, in Refs.\cite{log,mt} an asymptotic
expansion at the one-loop level, in the TeV range, 
was used of the more complete 
form $a U(q^2) + b_{\theta S} (\log q^2/M^2)_{\theta S} +
b_{RG} (\log q^2/M^2)_{RG} + c$. The numerical 
values of the constants $c$ in the various cases turned out to 
correspond systematically to a negative few percent 
(relative or absolute)
effect, with $a$ and $b$ in agreement with the theoretical 
Sudakov and RG
values. For LC energies, this decreased in all cases the 
overall logarithmic one loop effect, that was already in our 
philosophy under control, being the sum of two separately small effects.
>From this
fact we would be led to reinforce the conclusion that for LC energies
the complete one loop approximation should be satisfactory,
at least at the percent experimental level of accuracy.}. 
A full and detailed discussion on this
important point will appear soon in a forthcoming dedicated paper
\cite{top-2}.

\vspace{1cm}
{\bf Acknowledgments}: This work has been partially
supported by the European
Community grant  HPRN-CT-2000-00149.

\newpage

\appendix

\section{The special behaviour of the forward-backward asymmetries}

In this short appendix we investigate the origin
of the fact that the $\theta$-independent
contributions to $A_{FB,f}$ turn out to be small
in the high energy limit ($s>>M^2$).\par
In the separate production of chiral fermions ($f_{L}$ or $f_{R}$) 
a  $\theta$-independent correction $C^f_{L,R}$ does obviously
not modify the Born values of $A_{FB,f_{L,R}}$ as it gives the
same effect in the forward and in the backward domains.
However, since $C^e_{L}\neq C^e_{R}$,
$C^f_{L}\neq C^f_{R}$, and since the integrated Born
cross sections in the forward or in the backward domains 
$\sigma^{Born}_{F,B}(e_{L,R},f_{L,R})$ are non equal, it is
apparently not obvious that the forward-backward
asymmetry for unpolarized initial electrons and final fermions

\bqa
A_{FB,f}&&={\sigma_{F-B}(e_{L},f_{L})
+\sigma_{F-B}(e_{L},f_{R})
+\sigma_{F-B}(e_{R},f_{L})
+\sigma_{F-B}(e_{R},f_{R})
\over\sigma_{F+B}(e_{L},f_{L})
+\sigma_{F+B}(e_{L},f_{R})
+\sigma_{F+B}(e_{R},f_{L})
+\sigma_{F+B}(e_{R},f_{R})}\nonumber\\
&&=
[\sigma^{Born}_{F-B}(e_{L},f_{L})(1+C^e_{L}+C^f_{L})
+\sigma^{Born}_{F-B}(e_{L},f_{R})(1+C^e_{L}+C^f_{R})\nonumber\\
&&
+\sigma^{Born}_{F-B}(e_{R},f_{L})(1+C^e_{R}+C^f_{L})
+\sigma^{Born}_{F-B}(e_{R},f_{R})(1+C^e_{R}+C^f_{R})]\times
\nonumber\\
&&
[\sigma^{Born}_{F+B}(e_{L},f_{L})(1+C^e_{L}+C^f_{L})
+\sigma^{Born}_{F+B}(e_{L},f_{R})(1+C^e_{L}+C^f_{R})\nonumber\\
&&
+\sigma^{Born}_{F+B}(e_{R},f_{L})(1+C^e_{R}+C^f_{L})
+\sigma^{Born}_{F+B}(e_{R},f_{R})(1+C^e_{R}+C^f_{R})]^{-1}
\label{afbflr}
\eqa
\noindent
remains so close to its Born value.\par
The condition is that

\bqa
&&~~~~~[C^f_{L}-C^f_{R}][\sigma^{Born}_{F}(e_{L},f_{L})
(\sigma^{Born}_{B}(e_{L},f_{R})+\sigma^{Born}_{B}(e_{R},f_{R}))\nonumber\\
&&
-\sigma^{Born}_{B}(e_{L},f_{L})
(\sigma^{Born}_{F}(e_{L},f_{R})+\sigma^{Born}_{F}(e_{R},f_{R}))\nonumber\\
&&
-\sigma^{Born}_{F}(e_{L},f_{R})\sigma^{Born}_{B}(e_{R},f_{L})
+\sigma^{Born}_{B}(e_{L},f_{R})\sigma^{Born}_{F}(e_{R},f_{L})\nonumber\\
&&
+\sigma^{Born}_{F}(e_{R},f_{L})\sigma^{Born}_{B}(e_{R},f_{R})
-\sigma^{Born}_{B}(e_{R},f_{L})\sigma^{Born}_{F}(e_{R},f_{R})]\nonumber\\
&&
\simeq~~~ [C^e_{R}-C^e_{L}][\sigma^{Born}_{F}(e_{L},f_{L})
(\sigma^{Born}_{B}(e_{R},f_{L})+\sigma^{Born}_{B}(e_{R},f_{R}))\nonumber\\
&&
-\sigma^{Born}_{B}(e_{L},f_{L})
(\sigma^{Born}_{F}(e_{R},f_{L})+\sigma^{Born}_{F}(e_{R},f_{R}))\nonumber\\
&&
+\sigma^{Born}_{F}(e_{L},f_{R})\sigma^{Born}_{B}(e_{R},f_{L})
-\sigma^{Born}_{B}(e_{L},f_{R})\sigma^{Born}_{F}(e_{R},f_{L})\nonumber\\
&&
+\sigma^{Born}_{F}(e_{L},f_{R})\sigma^{Born}_{B}(e_{R},f_{R})
-\sigma^{Born}_{B}(e_{L},f_{R})\sigma^{Born}_{F}(e_{R},f_{R})]
\label{cond}\eqa
\noindent
and it is not trivially satisfied.\par
We have tried to analyse the contents of eq.(\ref{cond})
and to look for the origin of the cancellations which
appear in this expression or in the equivalent one,
Eqs.~(\ref{afbs}), given at the end of Appendix B,
which can be used in the case of
angular independent contributions:

\bq
A_{FB,f}= {3\over4}{U_{12}\over U_{11}}
\eq
\noindent
in which the photon and $Z$ exchange terms 
are explicitely written.\par
Simplifications arise when one uses the fact that
$s^2_W\simeq{1\over4}$ which makes the vector coupling of the $Z$
boson to $l^+l^-$ ($l=e,\mu,\tau$)
vanish (all the results obtained below would not be valid for
an arbitrary value of $s^2_W$).\par
We first consider in the $s^2_W\simeq{1\over4}$ approximation
the process $e^+e^-\to\mu^+\mu^-$, 
where the photon Born term is purely vector
and the $Z$ Born term purely axial.  One easily sees that
the one-loop corrections factorize
out in the same way ($1+2C^l_L+2C^l_R$)
in the numerator $U_{12}$ (only given by the
photon-$Z$ interference) and in
the denominator $U_{11}$ (only given by the squared photon and the
squared $Z$ terms), so that their total effect in $A_{FB,f}$
vanishes. So in practice these $\theta$-independent one-loop
corrections should be proportional to ($1-4s^2_W$) and indeed
very small.\par
The case of the processes $e^+e^-\to u\bar u$ and $e^+e^-\to d\bar d$
is less obvious. One still uses the fact that the
photon Born term is purely vector and that the \underline{initial}
$Z$ Born coupling to $e^+e^-$ is purely axial in the limit
$s^2_W={1\over4}$. In this limit, another essential ingredient
is the numerical value of

\bq 
{4\over3}A^{Born}_{FB,f}=\left({U_{12}\over U_{11}}\right)^{Born}
={3|Q_f|\over1-|Q_f|+5|Q_f|^2}
\eq
\noindent
which is close to $1$ for both up and down quarks,
i.e., ${18\over23}$ for $u$ and ${9\over11}$ for
$d$.\par
Including the angular independent corrections at first order
leads to:

\bq
{4\over3}A_{FB,f}=\left({3|Q_f|\over1-|Q_f|+5|Q_f|^2}\right)
\left({1+c^f_1+c^f_2\over1+c^f_1+c^f_3}\right)
\eq
\noindent
where
\bq
c^f_1=C^l_L+C^l_R+C^f_L+C^f_R
\eq
\bq
c^f_2=\left({1-|Q_f|\over|Q_f|}\right)
[|Q_f|(C^f_L-C^f_R)+{1\over3}(C^l_L-C^l_R)]
\eq
\bq
c^f_3=\left({3(1-|Q_f|)\over(1-|Q_f|+5|Q_f|^2}\right)
[|Q_f|(C^l_L-C^l_R)+{1\over3}(C^f_L-C^f_R)]
\eq

One sees now that, in addition to the term $c^f_1$ which would
factorize out like in the case $f=l$, there appear 
additional corrections $c^f_2$ and $c^f_3$ (which vanish for $f=l$).
However these additional corrections turn out to be both of the
same size, $c^f_2\simeq c^f_3$, and smaller than
$c^f_1$ in each of the cases $f=u$ and $f=d$. So at the end
the total correction to the Born value is again rather small.
One can trace the origin of the relation 
$c^f_2\simeq c^f_3<<c^f_1$ in the fact that,
using the notations of ref.\cite{Melles} for the
$\theta$-independent terms, the Left-handed
corrections $\simeq {3\over4}+{Y^2_L\over4}tan^2\theta_W$
are larger than the Right-handed ones 
$\simeq {Q^2_f\over4}tan^2\theta_W$ (this is the usual electroweak
feature), and also in the fact that
$3|Q_f|\simeq 1-|Q_f|+5|Q_f|^2$ (leading to
${4\over3}A^{Born}_{FB,f}\simeq 1$ for both $f=u,d$).\par
So in conclusion it appears that the angular independent
electroweak corrections to $A_{FB,f}$ turn out to be small
for "accidental" reasons related to the Left versus Right structure
of the electroweak multiplets and to the value $s^2_W\simeq{1\over4}$.
We do not see any deeper physical reason.

\newpage

\section{The general form of the polarized 
\lowercase{$e^+e^-\to f\bar f$}
cross section in the $Z$-peak subtracted representation}

The general expression of the $e^+e^-\to f\bar f$ cross section
can be written as

\bqa
{d\sigma_{f}\over dcos\theta}(P,P')&=&
{4\pi\over3}{\cal N}_fq^2\{{3\over8}
(1+cos^2\theta)[(1-PP')U_{11}+(P'-P)U_{21}]\nonumber\\
&&+{3\over4}cos\theta[(1-PP')
U_{12}+(P'-P)U_{22}]\}
\label{sigA}\eqa
\noindent
where
\bqa  
U_{11}=&&
{\alpha^2(0)Q^2_f\over q^4}[1+2
\tilde{\Delta}^{(lf)}_{\alpha}(q^2,\theta)]
\nonumber\\
&&+2[{\alpha(0)|Q_f|}]{q^2-M^2_Z\over
q^2((q^2-M^2_Z)^2+M^2_Z\Gamma^2_Z)}[{3\Gamma_l\over
M_Z}]^{1/2}[{3\Gamma_f\over {\cal N}_f M_Z}]^{1/2}
{\tilde{v}_l \tilde{v}_f\over
(1+\tilde{v}^2_l)^{1/2}(1+\tilde{v}^2_f)^{1/2}}\nonumber\\
&&\times[1+
\tilde{\Delta}^{(lf)}_{\alpha}(q^2,\theta) -R^{(lf)}(q^2,\theta)
-4s_lc_l
\{{1\over \tilde{v}_l}V^{(lf)}_{\gamma Z}(q^2,\theta)+{|Q_f|\over\tilde{v}_f}
V^{(lf)}_{Z\gamma}(q^2,\theta)\}]\nonumber\\ 
&&+{[{3\Gamma_l\over
M_Z}][{3\Gamma_f\over {\cal N}_f M_Z}]\over(q^2-M^2_Z)^2+M^2_Z\Gamma^2_Z}
\nonumber\\
&&\times[1-2R^{(lf)}(q^2,\theta)
-8s_lc_l\{{\tilde{v}_l\over1+\tilde{v}^2_l}V^{(lf)}_{\gamma
Z}(q^2,\theta)+{\tilde{v}_f|Q_f|\over(1+\tilde{v}^2_f)}
V^{(lf)}_{Z\gamma}(q^2,\theta)\}]
\label{U11pro}
\eqa
\bqa
U_{12}=&& 2[{\alpha(0)|Q_f|}]{q^2-M^2_Z\over
q^2((q^2-M^2_Z)^2+M^2_Z\Gamma^2_Z)}
[{3\Gamma_l\over M_Z}]^{1/2}[{3\Gamma_f\over {\cal N}_f
M_Z}]^{1/2}{1\over(1+\tilde{v}^2_l)^{1/2}(1+\tilde{v}^2_f)^{1/2}}
\nonumber\\
&&\times[1+
\tilde{\Delta}^{(lf)}_{\alpha}(q^2,\theta) -R^{(lf)}(q^2,\theta)]\nonumber\\
&&+{[{3\Gamma_l\over M_Z}][{3\Gamma_f\over {\cal N}_f
M_Z}]\over(q^2-M^2_Z)^2+M^2_Z\Gamma^2_Z}
[{4\tilde{v}_l \tilde{v}_f\over(1+\tilde{v}^2_l)(1+\tilde{v}^2_f)}]
\nonumber\\
&&\times[1-2R^{(lf)}(q^2,\theta)-4s_lc_l
\{{1\over \tilde{v}_l}V^{(lf)}_{\gamma Z}(q^2,\theta)+{|Q_f|\over\tilde{v}_f}
V^{(lf)}_{Z\gamma}(q^2,\theta)\}]
\label{U12pro}
\eqa
\bqa
U_{21}=&& 2[{\alpha(0)|Q_f|}]{q^2-M^2_Z\over
q^2((q^2-M^2_Z)^2+M^2_Z\Gamma^2_Z)}
[{3\Gamma_l\over
M_Z}]^{1/2}[{3\Gamma_f\over {\cal N}_f
M_Z}]^{1/2}{\tilde{v}_f\over(1+\tilde{v}^2_l)^{1/2}
(1+\tilde{v}^2_f)^{1/2}}\nonumber\\
&&\times[1+\tilde{\Delta}^{(lf)}_{\alpha}(q^2,\theta) -R^{(lf)}(q^2,\theta)
-{4s_lc_l|Q_f|\over\tilde{v}_f}V^{(lf)}_{Z\gamma}(q^2,\theta)]\nonumber\\
&&+{[{3\Gamma_l\over
M_Z}][{3\Gamma_f\over {\cal N}_f
M_Z}]\over(q^2-M^2_Z)^2+M^2_Z\Gamma^2_Z}
[{2\tilde{v}_l \over(1+\tilde{v}^2_l)}]\nonumber\\
&&\times[1-2R^{(lf)}(q^2,\theta)-4s_lc_l
\{{1\over \tilde{v}_l}V^{(lf)}_{\gamma
Z}(q^2,\theta)+{2\tilde{v}_f|Q_f|\over(1+\tilde{v}^2_f)}
V^{(lf)}_{Z\gamma}(q^2,\theta)\}]   
\label{U21pro}
\eqa
\bqa
U_{22}= && 2[{\alpha(0)|Q_f|}]{q^2-M^2_Z\over
q^2((q^2-M^2_Z)^2+M^2_Z\Gamma^2_Z)}
[{3\Gamma_l\over
M_Z}]^{1/2}[{3\Gamma_f\over {\cal N}_f
M_Z}]^{1/2}{\tilde{v}_l\over(1+\tilde{v}^2_l)^{1/2}
(1+\tilde{v}^2_f)^{1/2}}\nonumber\\
&&\times[1+\tilde{\Delta}^{(lf)}_{\alpha}(q^2,\theta) 
-R^{(lf)}(q^2,\theta)
-{4s_lc_l\over \tilde{v}_l}V^{(lf)}_{\gamma Z}(q^2)]\nonumber\\
&&+{[{3\Gamma_l\over
M_Z}][{3\Gamma_f\over {\cal N}_f
M_Z}]\over(q^2-M^2_Z)^2+M^2_Z\Gamma^2_Z}
[{2\tilde{v}_f \over(1+\tilde{v}^2_f)}]\nonumber\\
&&\times[1-2R^{(lf)}(q^2,\theta)-4s_lc_l
\{{2\tilde{v}_l\over(1+\tilde{v}^2_l)}
V^{(lf)}_{\gamma Z}(q^2,\theta)+{|Q_f|\over
\tilde{v}_f}V^{(lf)}_{Z\gamma}(q^2,\theta)\}] \label{U22pro}
\eqa 

Here $P,P'$ are the \underline{longitudinal} 
polarization degree of
the initial lepton and antilepton, and ${\cal N}_f$ is 
the colour factor for the $f\bar f$ channel which includes the
appropriate QCD corrections to the input.\par
>From this general expression one obtains the unpolarized integrated
cross section

\bq
\sigma_f=\int^{+1}_{-1} dcos\theta~{d\sigma_f\over dcos\theta}(0,0)
\label{sig}\eq

\noindent
the forward backward asymmetry

\bq
A_{FB,f}=\left(\int^{+1}_{0} dcos\theta~{d\sigma_f\over dcos\theta}(0,0)
-\int^{0}_{-1} dcos\theta~{d\sigma_f\over
dcos\theta}(0,0)\right)/\sigma_f
\label{afb}\eq

\noindent
and the longitudinal polarization asymmetry

\bq
A_{LR,f}=\left(\int^{+1}_{-1} dcos\theta~{d\sigma_f
\over dcos\theta}(-1,0)
-\int^{+1}_{-1} dcos\theta~{d\sigma_f\over
dcos\theta}(+1,0)\right)/2\sigma_f
\label{alr}\eq

Note that for $\theta$-independent contributions these integrals
simplify and allow to write

\bq
\sigma_f= {4\pi\over3}{\cal N}_f q^2 U_{11}
\label{sigs}\eq
\bq
A_{FB,f}= {3\over4}{U_{12}\over U_{11}}
\label{afbs}\eq
\bq
A_{LR,f}= {U_{21}\over U_{11}} \ .
\label{alrs}\eq

\clearpage
\newpage

\begin{figure}[p]
\[
\hspace{-0.5cm}\epsfig{file=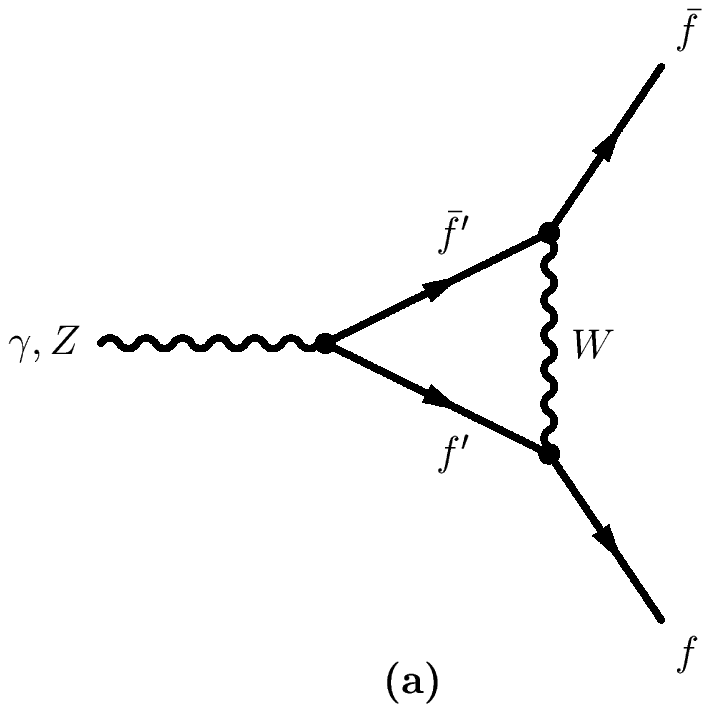,height=4cm}\hspace{1.cm}
\epsfig{file=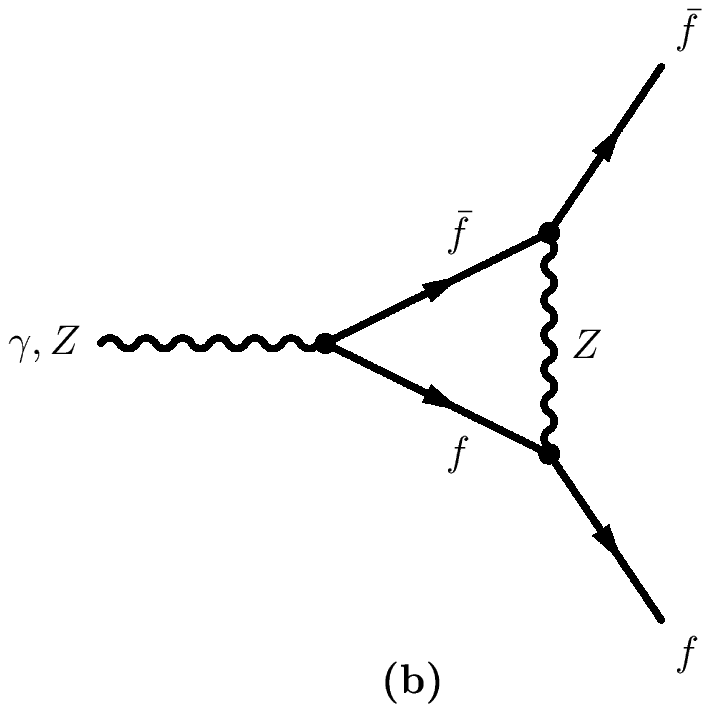,height=4cm}\hspace{1.cm}
\hspace{-0.5cm}\epsfig{file=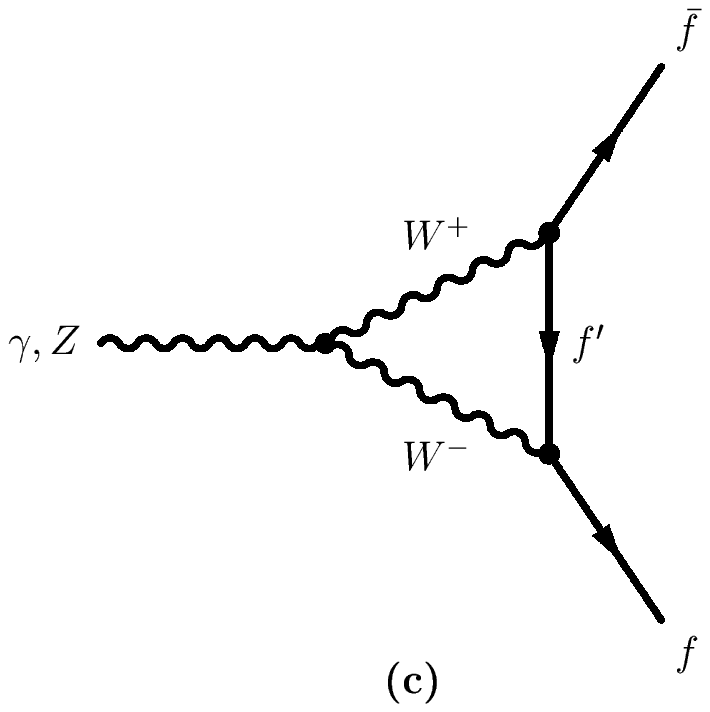,height=4cm}
\]
\vspace*{-0.5cm}
\[
\hspace{-0.5cm}\epsfig{file=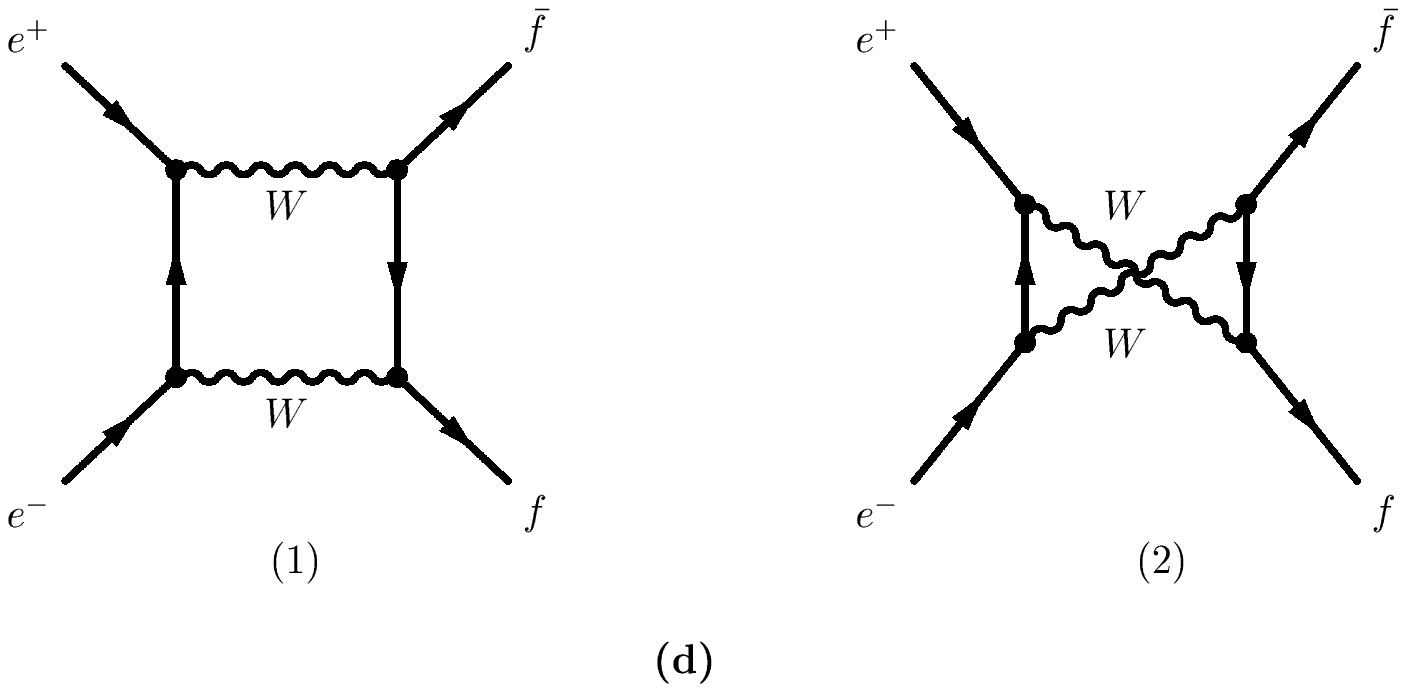,height=4cm}
\]
\vspace*{-0.5cm}
\[
\epsfig{file=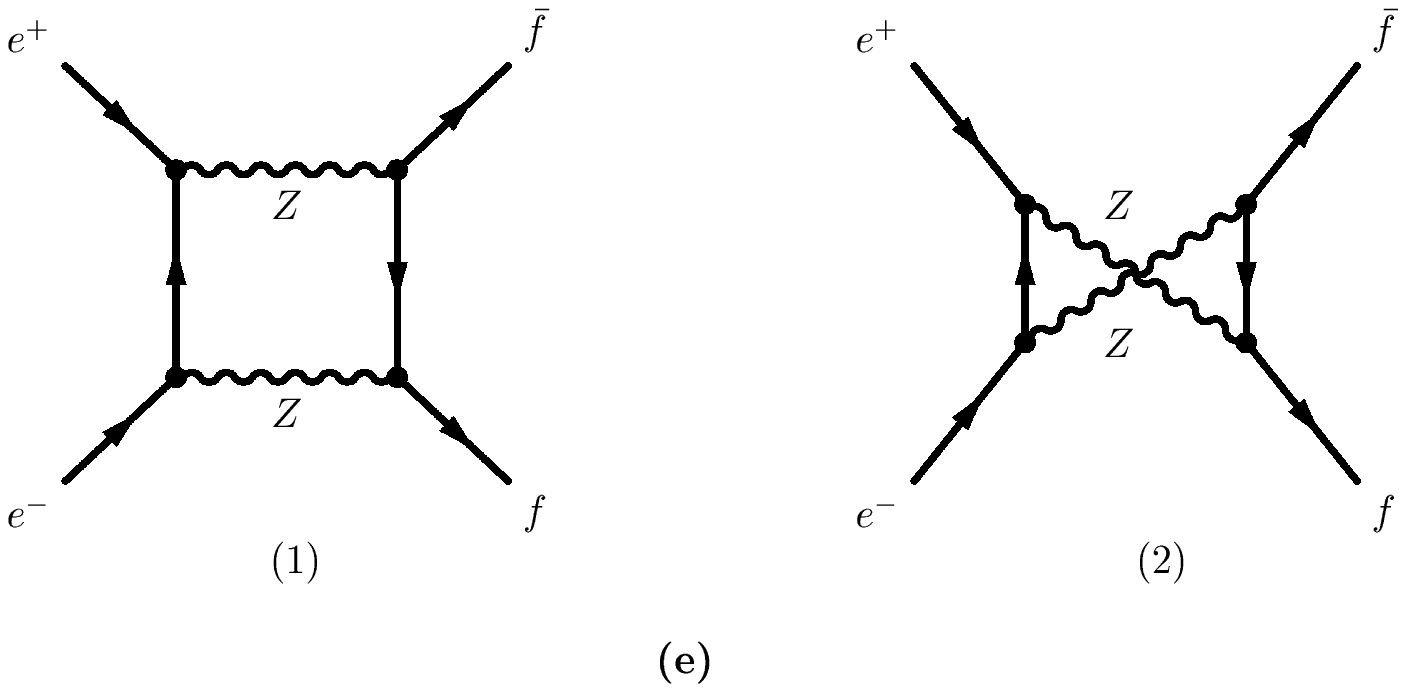,height=4cm}
\]
\vspace*{-0.5cm}
\[
\epsfig{file=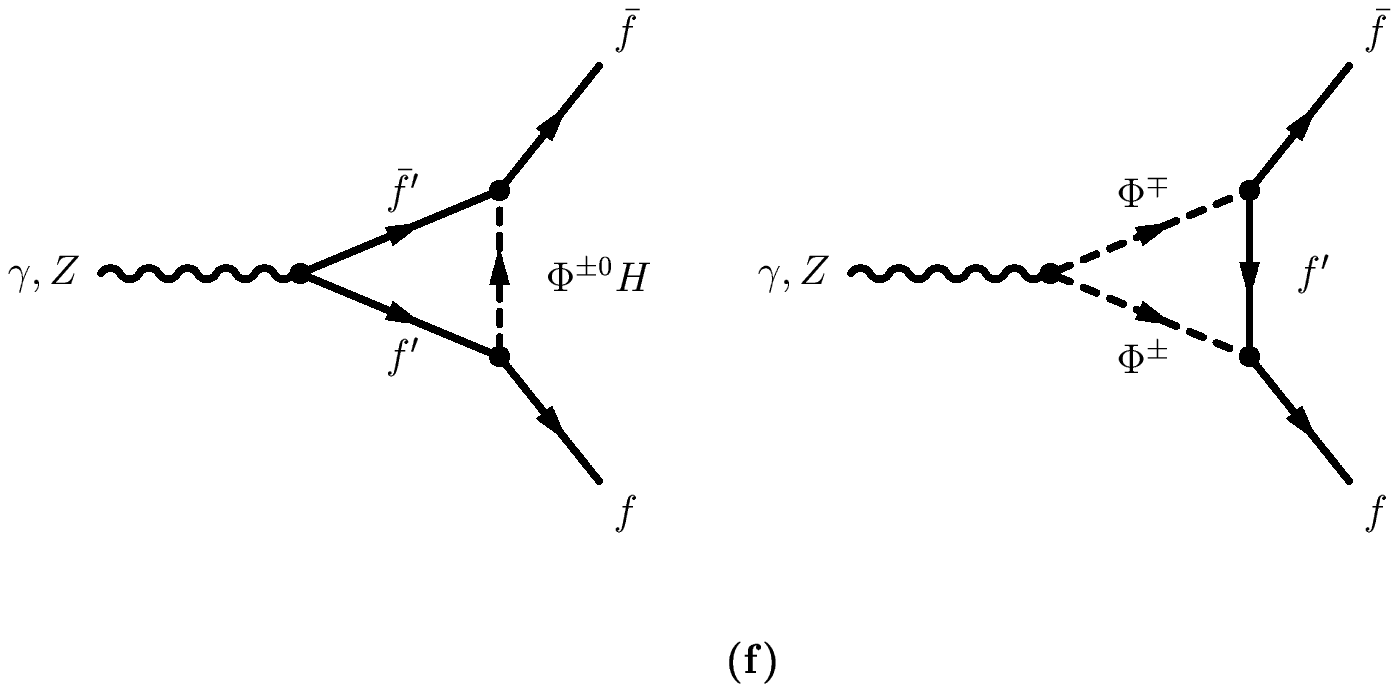,height=4cm}
\]
\vspace*{0cm}
\caption[1]{SM diagrams contributing in the asymptotic regime. Note
that for the $WW$ box contributions (d),
diagram (1) contributes for $I_{3f}=-{1\over2}$,
whereas diagram (2) contributes for $I_{3f}=+{1\over2}$; for the 
$ZZ$ box contributions (e), both diagrams contribute.}
\label{fig1}
\end{figure}

\begin{figure}[htb]
\vspace*{2cm}
\[
\epsfig{file=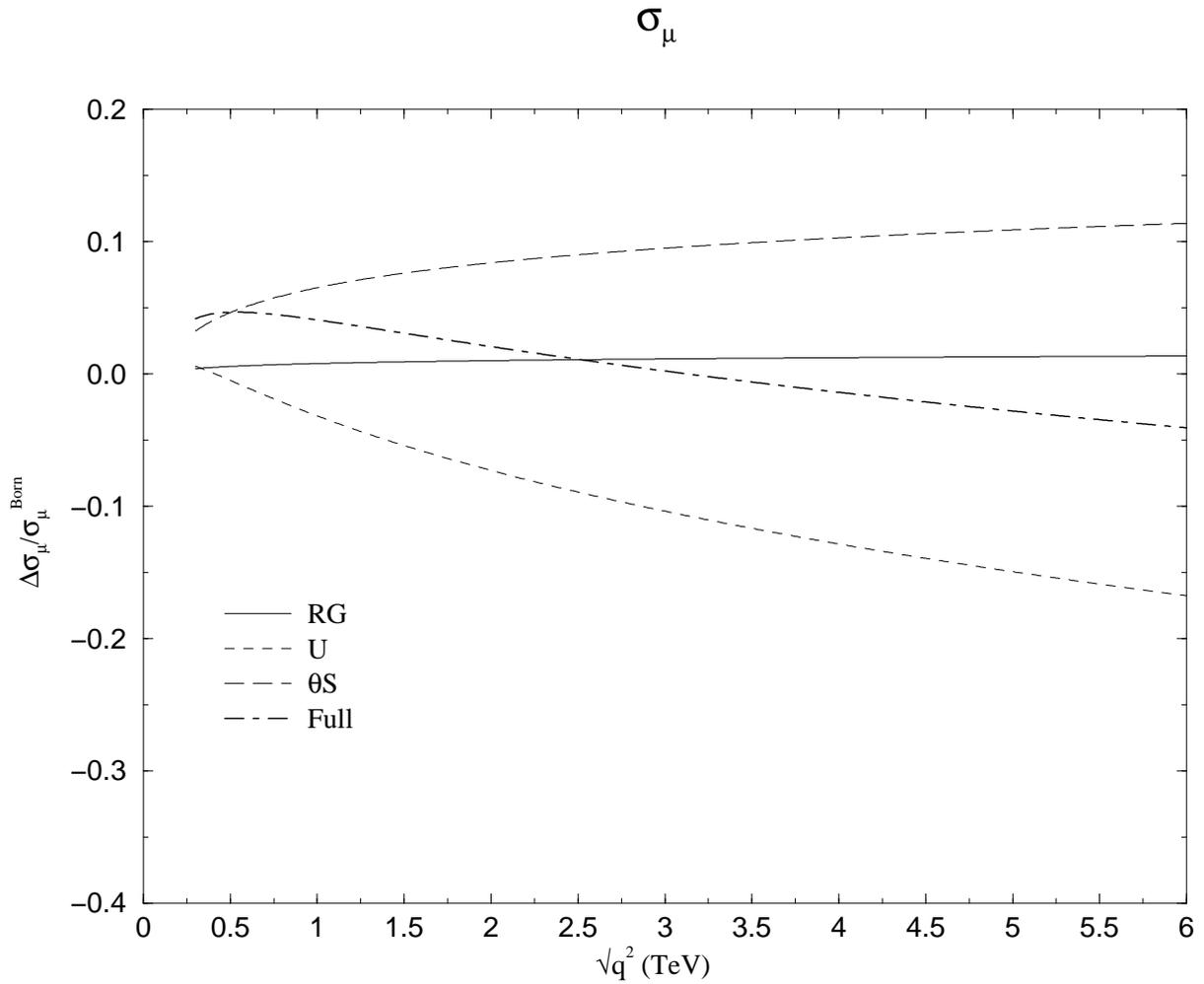,height=16cm,angle=-90}
\]
\vspace*{1cm}
\caption[1]{
Separate asymptotic contributions to $\sigma_\mu$ as functions of the energy.
The  solid line (RG) is the linear Renormalization Group logarithm. The 
dotted line (U) is the $\theta$ independent term proportional to the
combination $3\log\frac{q^2}{M_Z^2}-\log^2\frac{q^2}{M_Z^2}$. The dashed line
($\theta S$) 
is the angular dependent linear logarithm. Finally, the thick dot-dashed
line (Full) is the sum of the three contributions. The same captions applies 
to all the following figures showing the effects in all the other considered
observables.}
\label{figsigmamu}
\end{figure}
\newpage

\begin{figure}[htb]
\vspace*{2cm}
\[
\epsfig{file=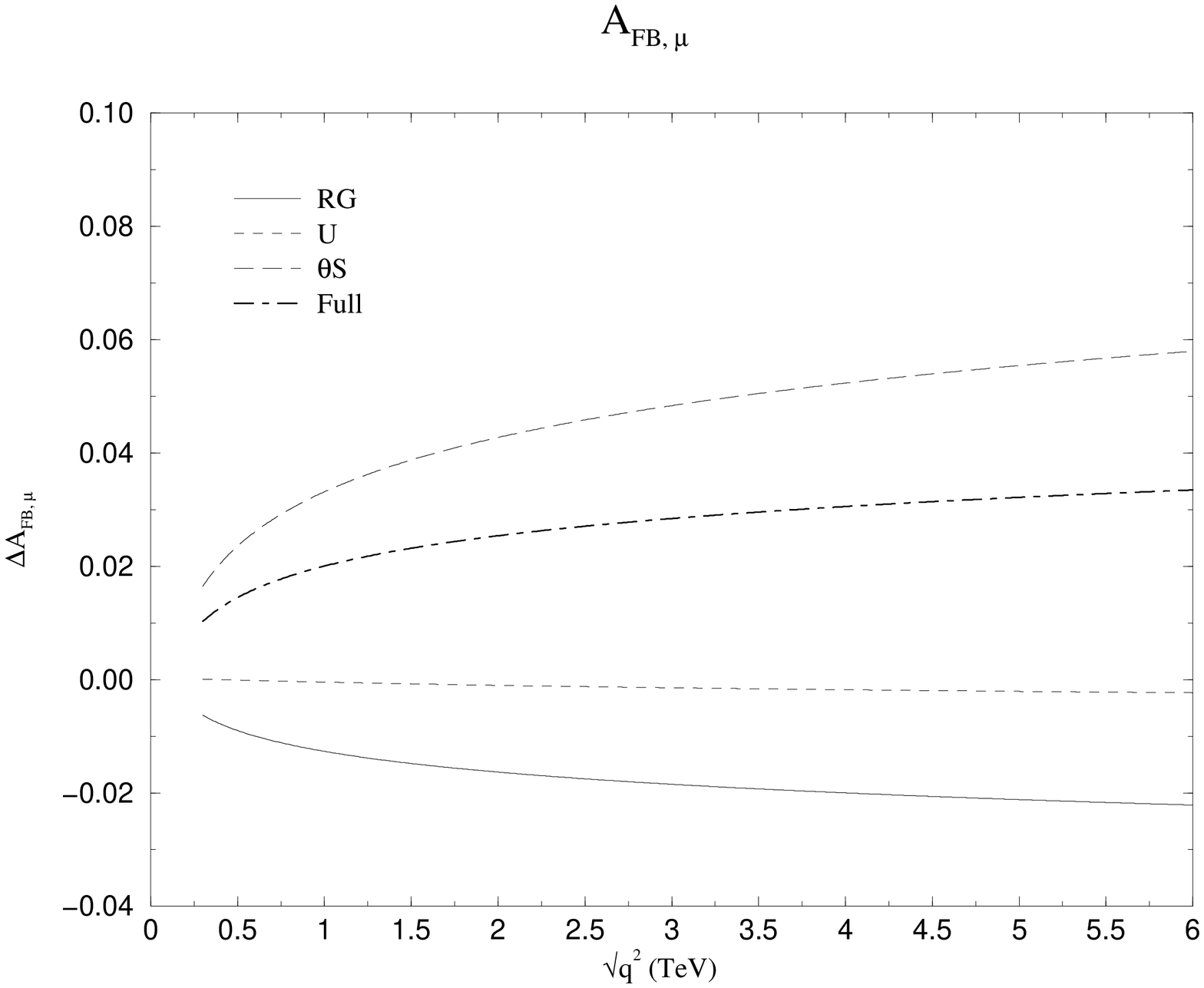,height=16cm,angle=-90}
\]
\vspace*{1cm}
\caption[1]{
Separate asymptotic contributions to $A_{FB,\mu}$ 
as functions of the energy. Same captions as in Fig.2.}
\label{figafbmu}
\end{figure}
\newpage

\begin{figure}[htb]
\vspace*{2cm}
\[
\epsfig{file=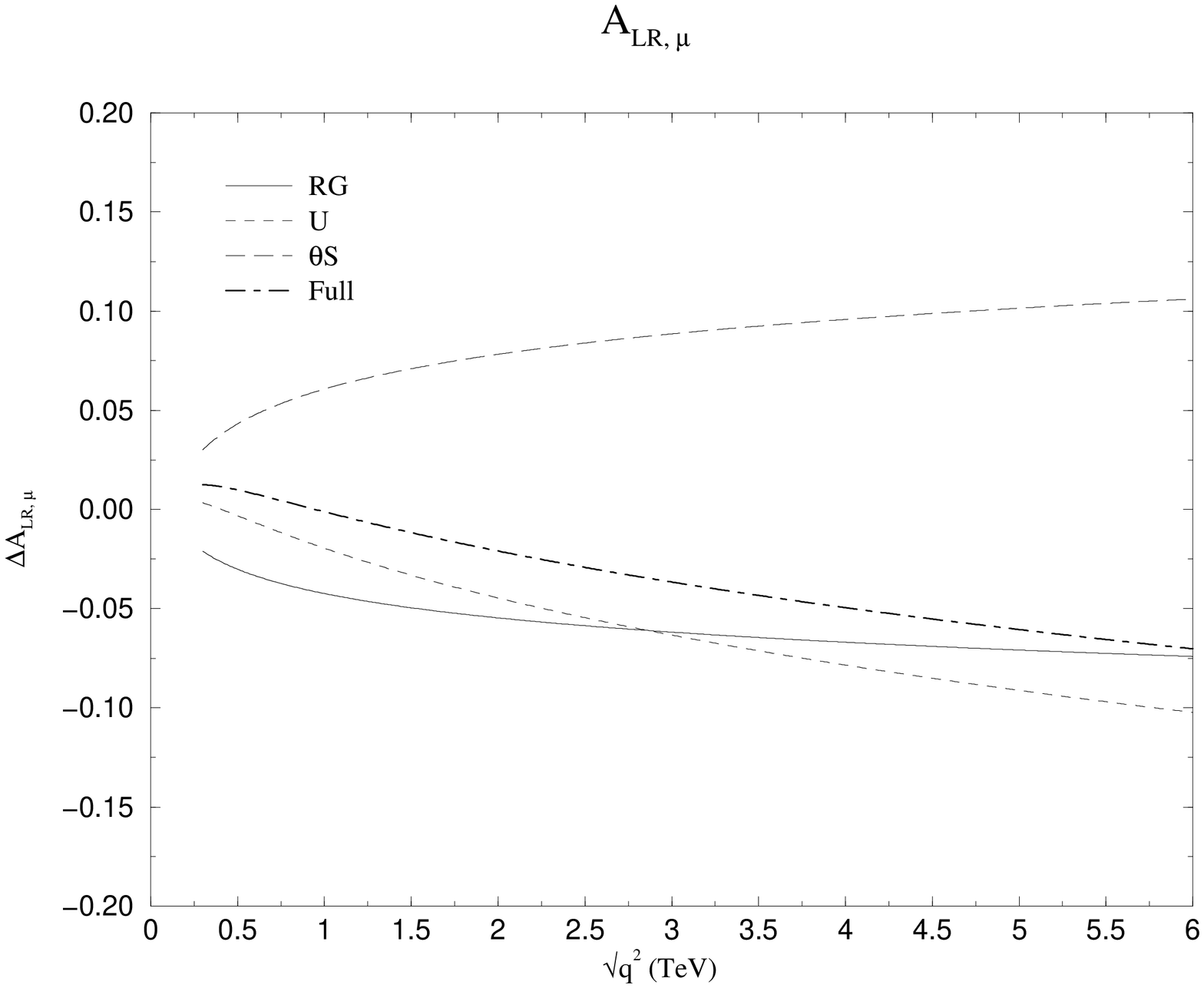,height=16cm,angle=-90}
\]
\vspace*{1cm}
\caption[1]{
Separate asymptotic contributions to $A_{LR,\mu}$ 
as functions of the energy. Same captions as in Fig.2.}
\label{figalrmu}
\end{figure}
\newpage

\begin{figure}[htb]
\vspace*{2cm}
\[
\epsfig{file=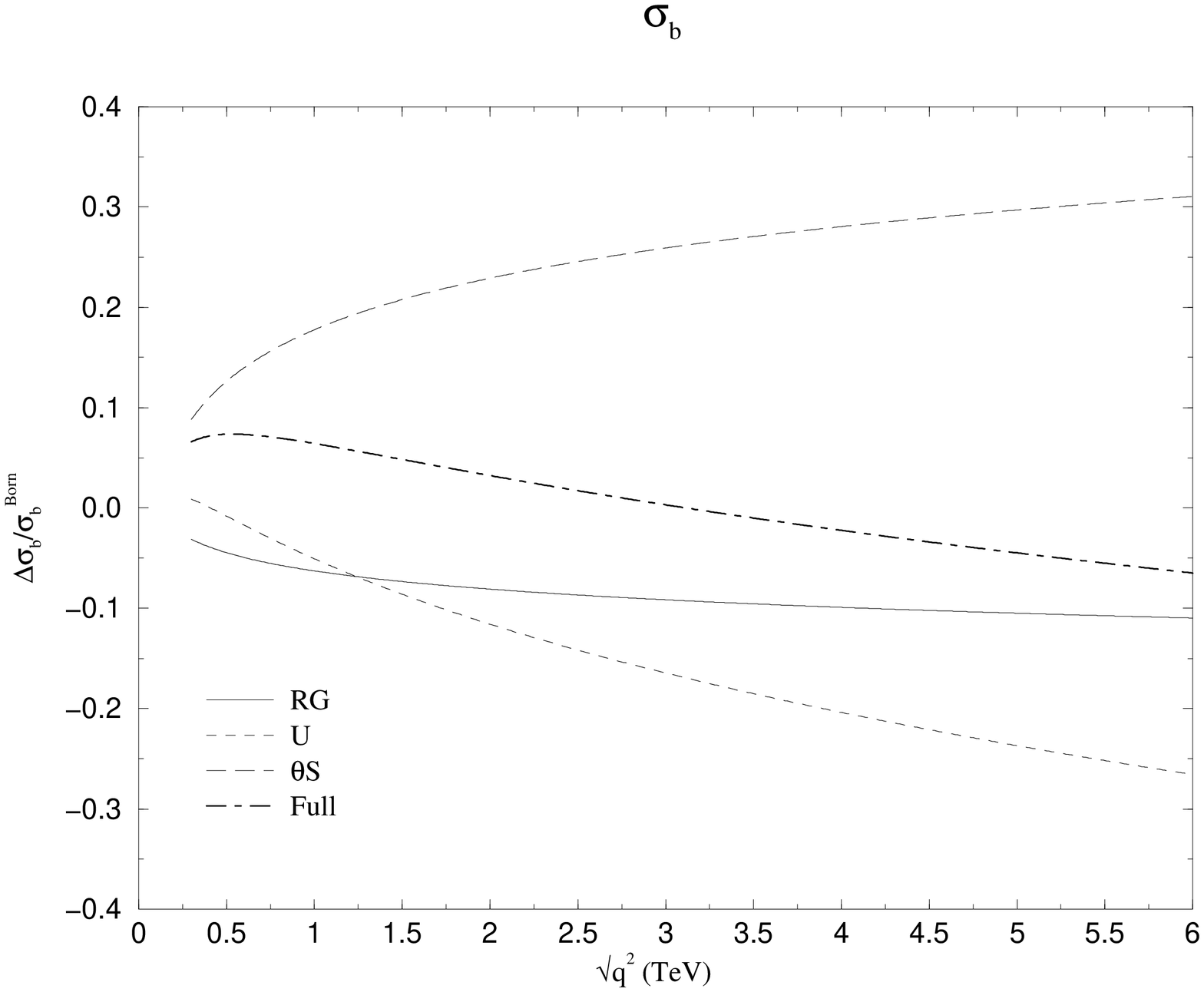,height=16cm,angle=-90}
\]
\vspace*{1cm}
\caption[1]{
Separate asymptotic contributions to $\sigma_b$ 
as functions of the energy. Same captions as in Fig.2.}
\label{figsigmab}
\end{figure}
\newpage

\begin{figure}[htb]
\vspace*{2cm}
\[
\epsfig{file=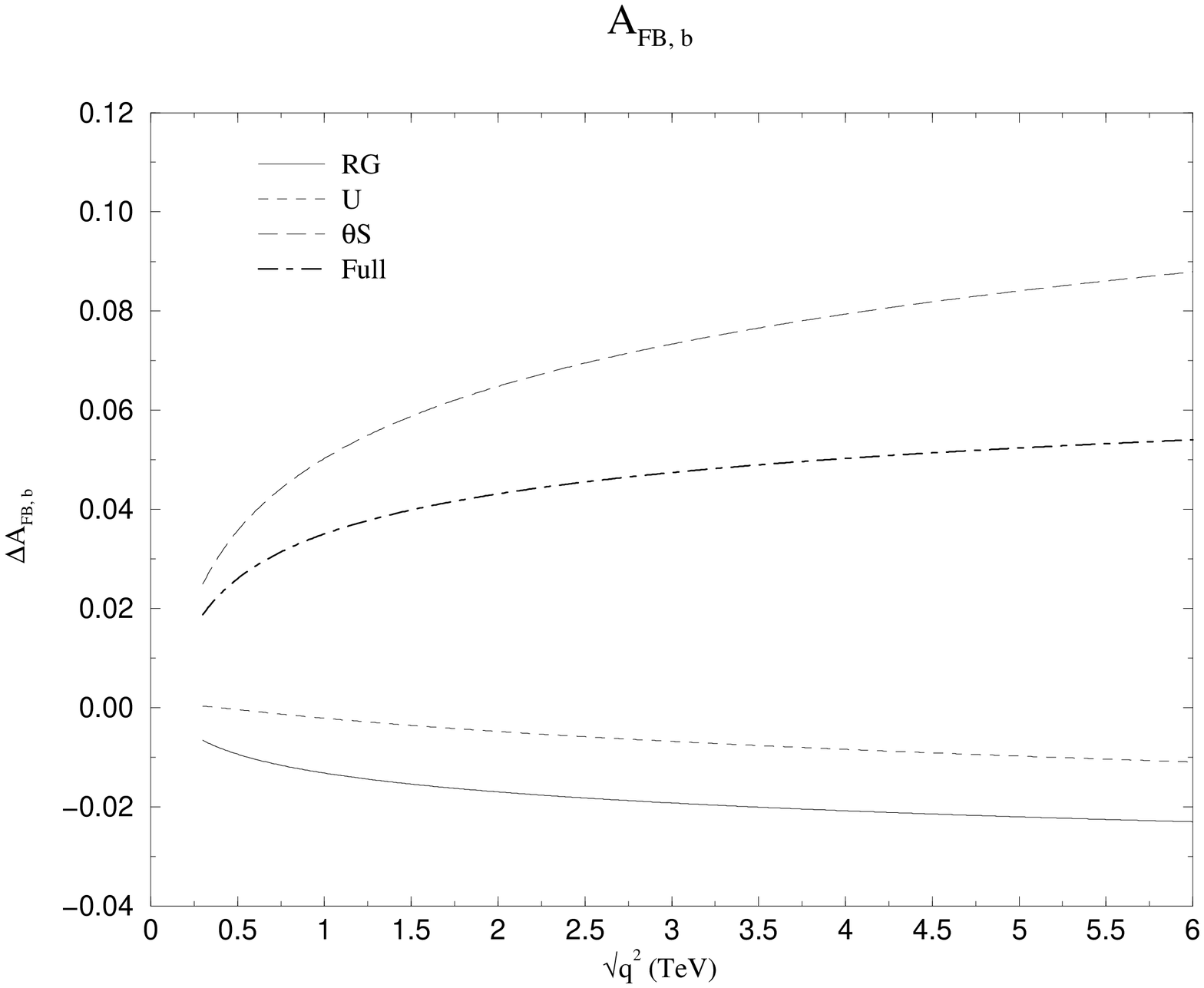,height=16cm,angle=-90}
\]
\vspace*{1cm}
\caption[1]{
Separate asymptotic contributions to $A_{FB,b}$ 
as functions of the energy. Same captions as in Fig.2.}
\label{figafbb}
\end{figure}
\newpage

\begin{figure}[htb]
\vspace*{2cm}
\[
\epsfig{file=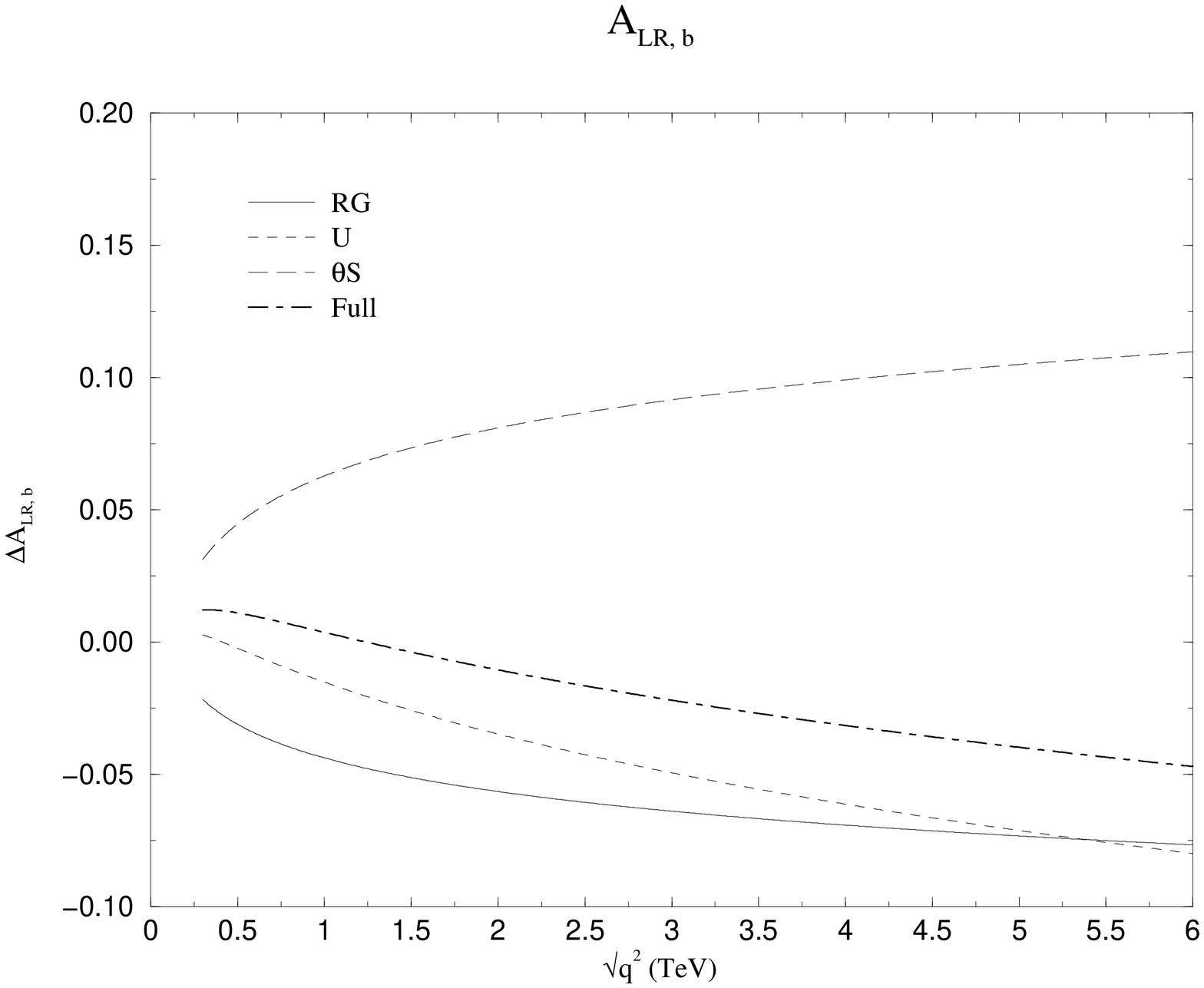,height=16cm,angle=-90}
\]
\vspace*{1cm}
\caption[1]{
Separate asymptotic contributions to $A_{LR,b}$ 
as functions of the energy. Same captions as in Fig.2.}
\label{figalrb}
\end{figure}
\newpage

\begin{figure}[htb]
\vspace*{2cm}
\[
\epsfig{file=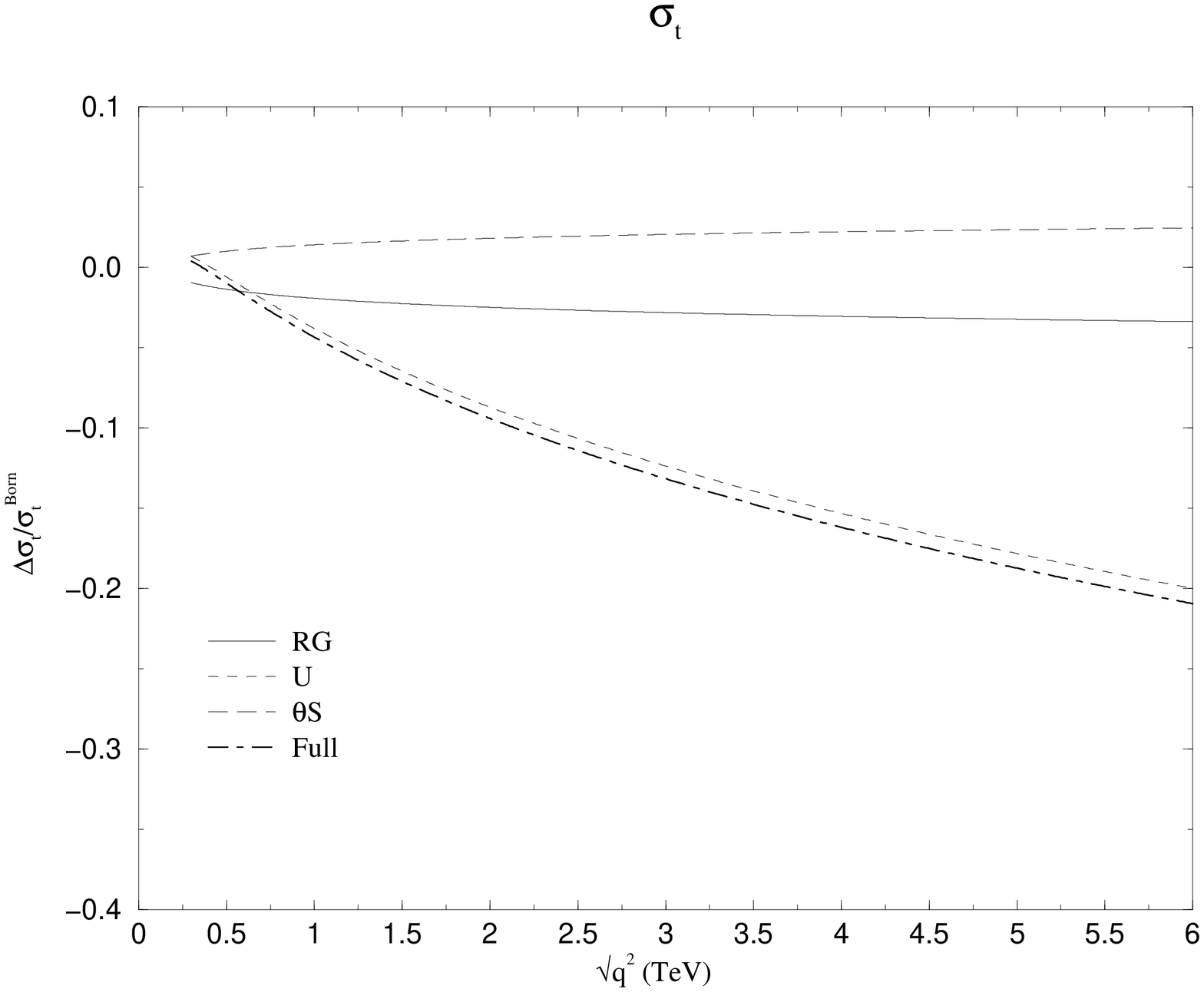,height=16cm,angle=-90}
\]
\vspace*{1cm}
\caption[1]{
Separate asymptotic contributions to $\sigma_t$ 
as functions of the energy. Same captions as in Fig.2.}
\label{figsigmat}
\end{figure}
\newpage

\begin{figure}[htb]
\vspace*{2cm}
\[
\epsfig{file=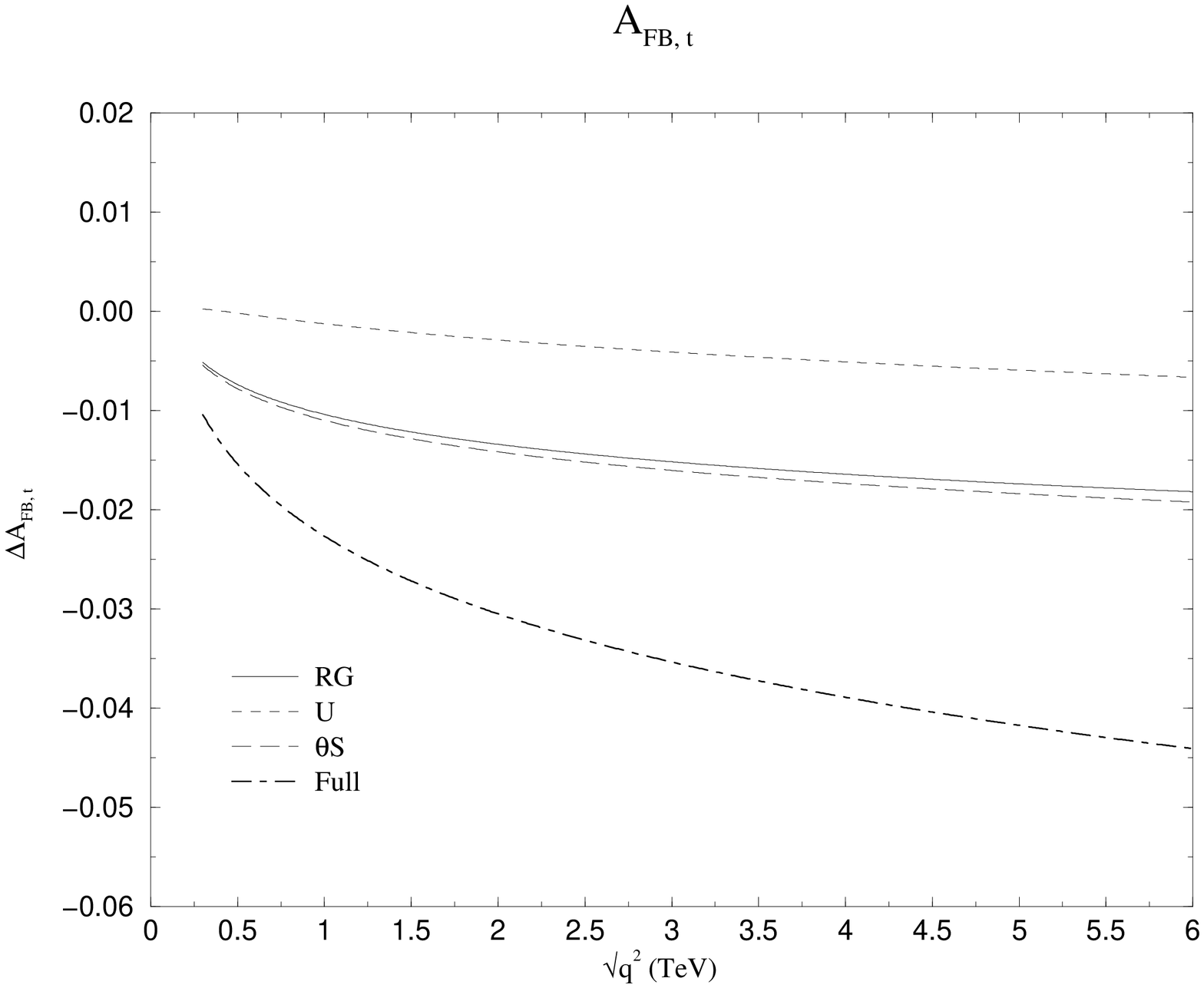,height=16cm,angle=-90}
\]
\vspace*{1cm}
\caption[1]{
Separate asymptotic contributions to $A_{FB,t}$ 
as functions of the energy. Same captions as in Fig.2.}
\label{figafbt}
\end{figure}
\newpage

\begin{figure}[htb]
\vspace*{2cm}
\[
\epsfig{file=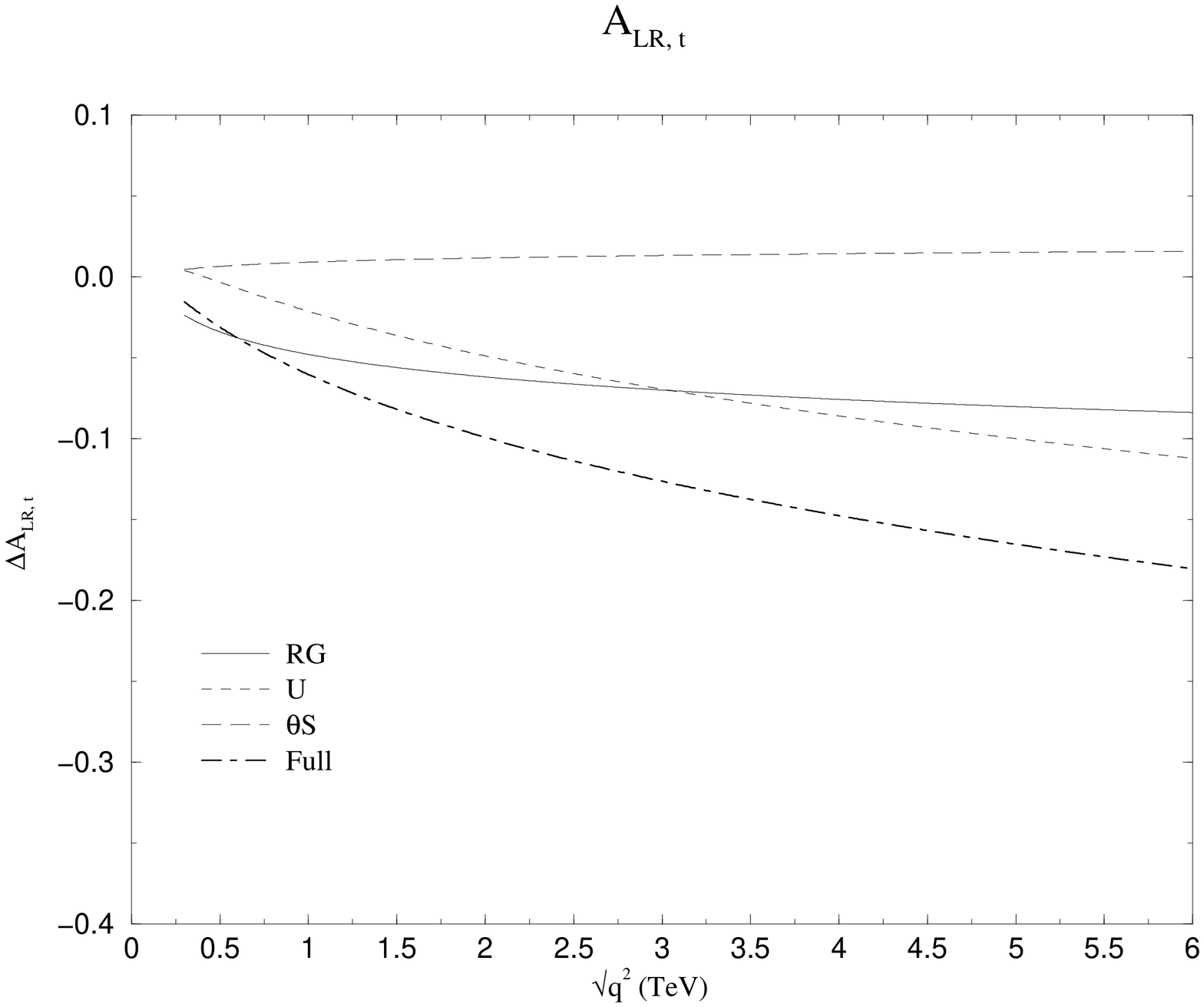,height=16cm,angle=-90}
\]
\vspace*{1cm}
\caption[1]{
Separate asymptotic contributions to $A_{LR,t}$ 
as functions of the energy. Same captions as in Fig.2.}
\label{figalrt}
\end{figure}
\newpage

\begin{figure}[htb]
\vspace*{2cm}
\[
\epsfig{file=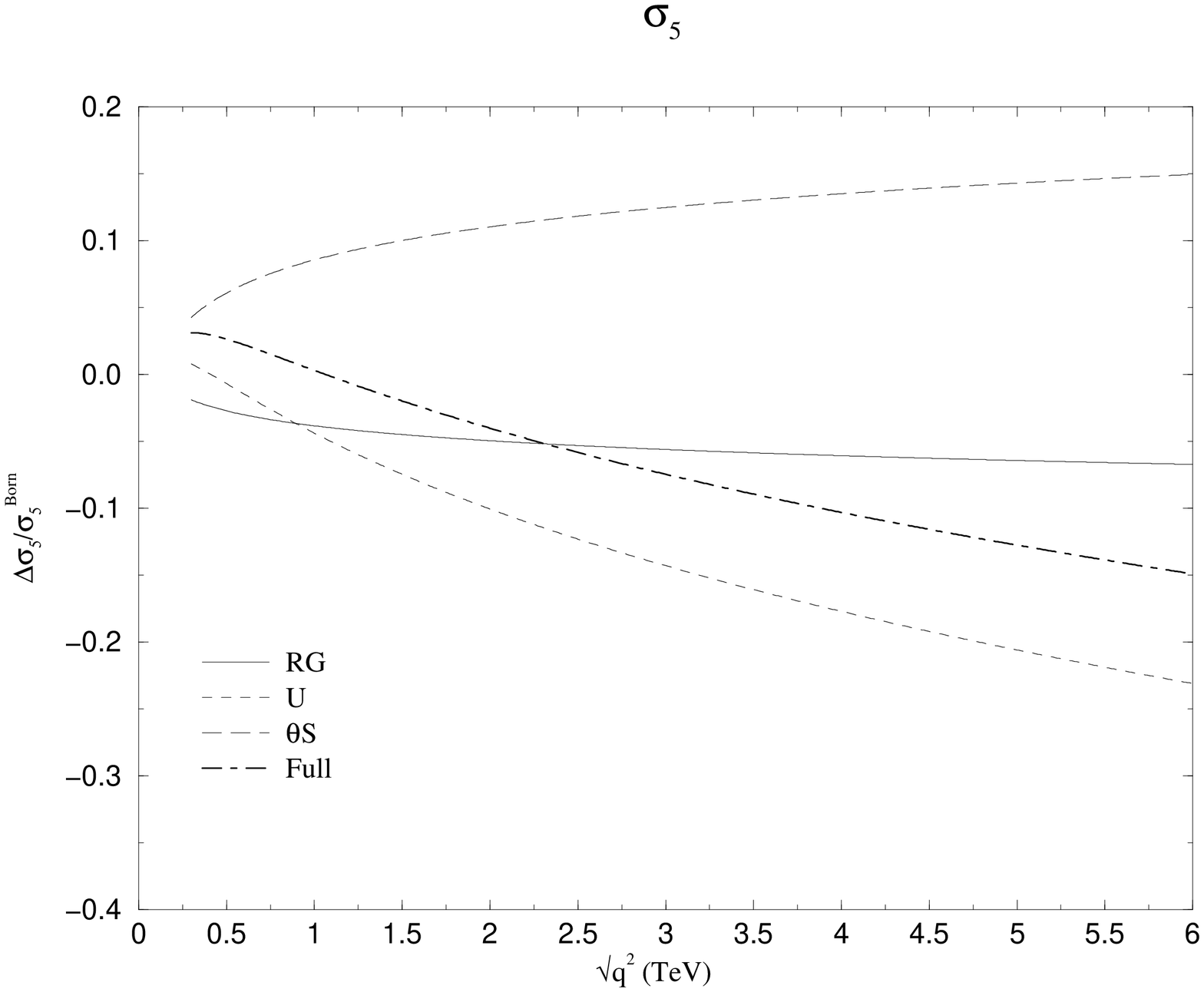,height=16cm,angle=-90}
\]
\vspace*{1cm}
\caption[1]{
Separate asymptotic contributions to $\sigma_5$ 
as functions of the energy. Same captions as in Fig.2.}
\label{figsigma5}
\end{figure}
\newpage

\begin{figure}[htb]
\vspace*{2cm}
\[
\epsfig{file=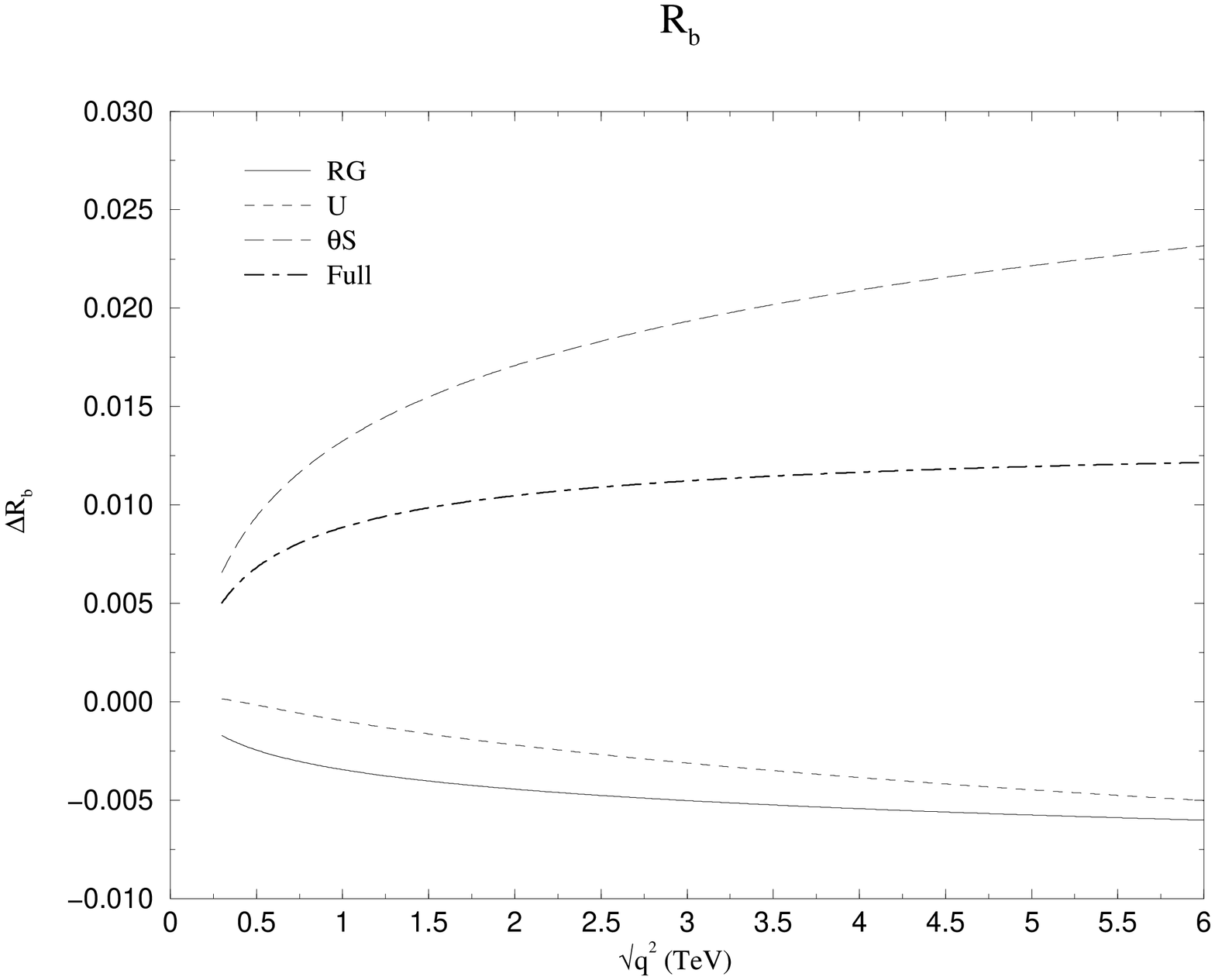,height=16cm,angle=-90}
\]
\vspace*{1cm}
\caption[1]{
Separate asymptotic contributions to $R_b$ 
as functions of the energy. Same captions as in Fig.2.}
\label{figrb}
\end{figure}
\newpage

\begin{figure}[htb]
\vspace*{2cm}
\[
\epsfig{file=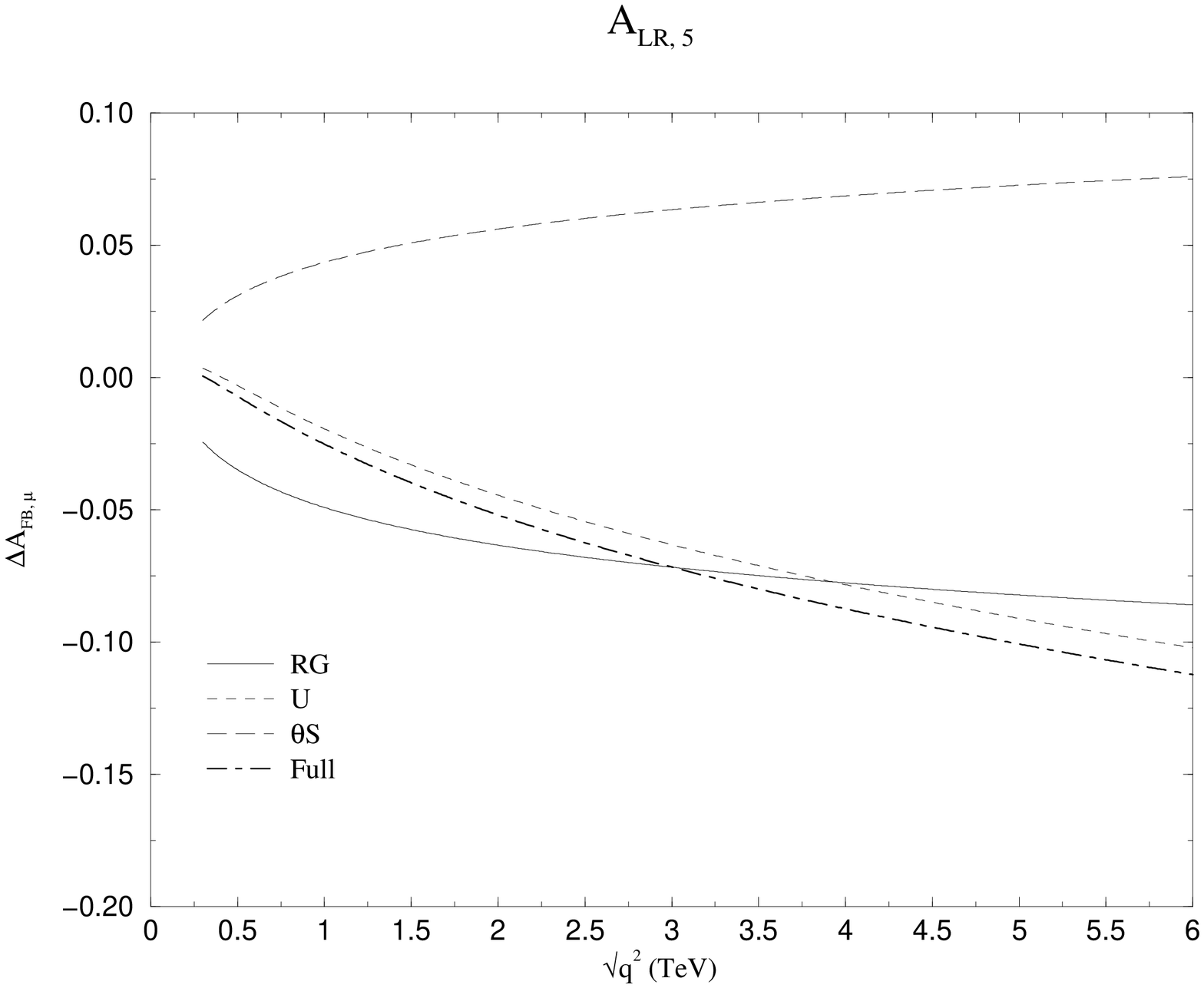,height=16cm,angle=-90}
\]
\vspace*{1cm}
\caption[1]{
Separate asymptotic contributions to $A_{LR,5}$ 
as functions of the energy. Same captions as in Fig.2.}
\label{figalr5}
\end{figure}
\newpage

\end{document}